\documentclass[preprint,12pt]{elsarticle}
\usepackage{amssymb}
\usepackage{amsthm}
\usepackage{float}
\usepackage{xcolor}
\usepackage[caption = false]{subfig}
\usepackage{graphicx}
\usepackage{varwidth}
\usepackage{colortbl}
\usepackage[hidelinks]{hyperref}
\usepackage{stix}
\usepackage{cleveref}
\usepackage{siunitx}
\usepackage{amsbsy}

\DeclareSIUnit{\molar}{M}

\usepackage{epstopdf}
\usepackage[english]{babel}
\usepackage{svg}
\usepackage{pgfplots}
\usepackage[T1]{fontenc}
\journal{Journal of Molecular Liquids}
\begin{document}

\begin{frontmatter}



\title{Role of ionic surfactant in magnetic dynamics of self-assembled dispersions of nanoplatelets}


%\author[inst1]{Author One}
%
%\affiliation[inst1]{organization={Department One},%Department and Organization
%            addressline={Address One}, 
%            city={City One},
%            postcode={00000}, 
%            state={State One},
%            country={Country One}}

\author[inst1]{Hajnalka N\'{a}dasi\corref{cor1}}
\author[inst2]{Melvin K\"{u}ster}
\author[inst3]{Alenka Mertelj}
\author[inst3]{Nerea Sebasti\'{a}n}
\author[inst3,inst4]{Patricija Hribar Bo\u{s}tjan\u{c}i\u{c}}
\author[inst5]{Darja Lisjak}
\author[inst2]{Thilo Viereck}
\author[inst6]{Margaret Rosenberg}
\author[inst8,inst9]{Alexey O. Ivanov}
\author[inst6,inst7]{Sofia S. Kantorovich}
\author[inst1]{Alexey Eremin}
\author[inst2]{Frank Ludwig\corref{cor1}}

\cortext[cor1]{Corresponding authors. E-mail adresses: \newline
\href{hajnalka.nadasi@ovgu.de}{hajnalka.nadasi@ovgu.de} (H. Nádasi)\newline \href{f.ludwig@tu-braunschweig.de}{f.ludwig@tu-braunschweig.de} (F. Ludwig).}

%\cortext[cor1]{Corresponding authors. E-mail adresses:\newline \href{f.ludwig@tu-braunschweig.de}{f.ludwig@tu-braunschweig.de} (F. Ludwig),\newline
%\href{hajnalka.nadasi@ovgu.de}{hajnalka.nadasi@ovgu.de} (H. Nádasi).}

\affiliation[inst1]{organization={Institute of Physics, Otto von Guericke University},%Department and Organization
            addressline={Universitätsplatz 2}, 
            city={Magdeburg},
            postcode={39106},
            country={Germany}}

\affiliation[inst2]{organization={Institute of Electrical Measurement Science and Fundamental Electrical Engineering (EMG) and Laboratory for Emerging Nanometrology (LENA), TU Braunschweig},%Department and Organization
            addressline={Hans-Sommer-Str. 66}, 
            city={Braunschweig},
            postcode={38106},
            country={Germany}}

\affiliation[inst3]{organization={Department of Complex Matter, Jožef Stefan Institute},%Department and Organization
            addressline={Jamova cesta 39}, 
            city={Ljubljana},
            postcode={SI-1000},
            country={Slovenia}}

\affiliation[inst4]{organization={Jožef Stefan International Postgraduate School},%Department and Organization
            addressline={Jamova cesta 39}, 
            city={Ljubljana},
            postcode={SI-1000},
            country={Slovenia}}

\affiliation[inst5]{organization={Department for Materials Synthesis, Jožef Stefan Institute},
            addressline={Jamova cesta 39}, 
            city={Ljubljana},
            postcode={SI-1000},
            country={Slovenia}}

\affiliation[inst6]{
            organization={Faculty of Physics, University of Vienna},
            addressline={Boltzmanngasse 5},
            city={Vienna},
            postcode={1090},
            state={},
            country={Austria}}
            
\affiliation[inst7]{
            organization={Research Platform ``Mathematics-Magnetism-Materials'', University of Vienna},
            addressline={Oskar-Morgenstern-Platz 1},
            city={Vienna},
            postcode={1090},
            country={Austria}}     

\affiliation[inst8]{
            organization={Ural Federal University},
            addressline={51 Lenin Ave.},
            city={Ekaterinburg},
            country={Russian Federation }}     

\affiliation[inst9]{
            organization={ M.N. Mikheev Institute of Metal Physics UB RAS},
            addressline={18 S. Kovalevskaya str.},
            city={Ekaterinburg},
            country={Russian Federation }}      

\begin{abstract}
In complex colloidal systems, interparticle interactions strongly affect the dynamics of the constituting particles. A study of the dynamical response also provides invaluable information on the character of those interactions. Here we demonstrate how tuning the electrostatic interactions by an ionic surfactant in dispersions of magnetic nanoplatelets leads to developing new dynamic modes in magnetic response spectra. The collective modes can be induced or suppressed by either varying the concentration ratio of the magnetic nanoplatelets (MP) to the surfactant or increasing the MP concentration reflecting the nanoscale characteristics of this fluid magnet. 
\end{abstract}




\end{frontmatter}


%\section{TO-Do's}

\section{Introduction}
\label{sec:sample1}

Ferrofluids are one of the most spectacular examples of complex fluids where nanoscale properties can be transformed to the macroscopic level \cite{rosensweig2013ferrohydrodynamics, BlumsCebersMaiorov2010, rinaldifranklinzahncader}. Understanding the nanoscale dynamics in such systems allows accurate manipulation and fine-tuning of their macroscopic properties \cite{LudwigRemmer2022,Hess:2020bj,Rupnik:2015bi, Lee:2018dl}. 
These dynamics are governed by two key attributes of the system: the electrostatic interactions, which ensure the colloidal stability of the magnetic nanoparticles in the liquid dispersion medium, and the magnetic dipole moment of the nanoparticles. Understand the interplay between these two properties is critical to the tailoring of the ferrofluids’s microstructure, which can then give rise to the desired macroscopic properties.

Surfactants stabilise colloidal suspensions by physical or chemical adsorption on the suspended particles' surfaces. Depending on the solvent affinity of the surfactant’s head either they form a single layer around the particles adsorbed by their solvophobic head(s) while dispersing them in the medium by the solvophilic tail, or they organise into a double layer, whereby the overlapping solvophobic tails are enclosed by the solvophilic heads. The key to successfully integrating the particles in the dispersing medium is to match the dielectric properties of the surfactant and the solvent. However, the suspension's longevity depends on the effective prevention of aggregation \cite{russel_saville_schowalter_1989, Hoeven.1992}. In the case of magnetic fluids, it is also essential to counteract the colloidal self-assembly caused by dipolar interactions, which can be significantly enhanced by the magnetic field. While reversible agglomeration is desirable in specific use cases, such as magnetooptical effects \cite{0m8, Mendelev:2004ve, Nadasi:2019cs,NadasiBook}, extensive irreversible aggregation of dipolar nanoparticles is generally undesirable due to the loss of colloidal stability.

Ferrimagnetic scandium-doped barium hexaferrite platelets (Sc-BaHF) are (shape-)~anisotropic particles with their magnetisation perpendicular to the basal plane \cite{Lisjak.2016, Haehsler.2020}.

When such platelets are dispersed into a nematic liquid crystal matrix (5CB (4-Cyano-4'-pentylbiphenyl), E7), the anchoring of the nematic director at the particle surfaces may stabilise the colloidal ferromagnetic nematic order as demonstrated in \cite{Mertelj:2014kv, Mertelj.2019, Rupnik:2017wd}. It is the first example of the long-anticipated ferromagnetic nematic, as predicted by de Gennes \cite{deGennes:1995vg}. Ferromagnetic nematic order can also be realised in the pure colloidal liquid crystal, where the nanoplatelets are dispersed in an isotropic fluid, e.g. 1-butanol \cite{Shuai:2016fd, Bostjancic.2022, Gregorin.2022}.

The high magnetic moment of ferrimagnetic Sc-BaHF platelets requires electrostatic stabilisation against aggregation, which can be achieved by an ionic surfactant such as dodecylbenzenesulfonic acid (DBSA) \cite{Lisjak.2011}.

In our previous paper \cite{Kuester.2022} we investigated the dynamic magnetic response of a dilution series of Sc-BaHF ferrofluids.

The dispersion medium 1-butanol contained \qty{4.5}{\milli\molar} surfactant (DBSA) in the stock suspension. To ensure the stability of the ferrofluid, it was diluted by the stock solution with the same concentration of DBSA during preparation. In an oscillating magnetic field, we observed collective modes of the as-prepared ferrofluid series. The collective modes were well resolved in the AC susceptibility (ACS) spectra of the ferrofluids with low magnetic particle concentration.
A possible explanation for the appearance of those modes is that as we dilute the stock suspension by \qty{4.5}{\milli\molar} DBSA solution, the degree of concentration-change is different for the MPs compared to that of the DBSA.

The adsorption-desorption equilibrium of the surfactant at the MP surface depends not only on the DBSA but also on the MP concentration ($c_\mathrm{MP}$). Bo\v{s}tjan\v{c}i\v{c} et al. \cite{Bostjancic.2019} investigated suspensions of Sc-BaHF platelets in the concentration range of $c_\mathrm{MP}=\qtyrange[range-phrase=~-~]{5}{30}{\gram\per\liter}$. The ratio of the adsorbed DBSA to the total amount of DBSA in the suspension increases with increasing platelet concentration. The dissolved molecules in the dispersing medium are partly dissociated. With increasing concentration of this ionic form, the ionic strength increases, whereby the Debye screening length of the platelets decreases. The repulsive electrostatic interactions also depend on the effective charge of the platelets, which was found to almost independent of the DBSA concentration ($c_\mathrm{DBSA}$).

\begin{figure}[hbt]
\centering
  \includegraphics[width=0.7\columnwidth ]{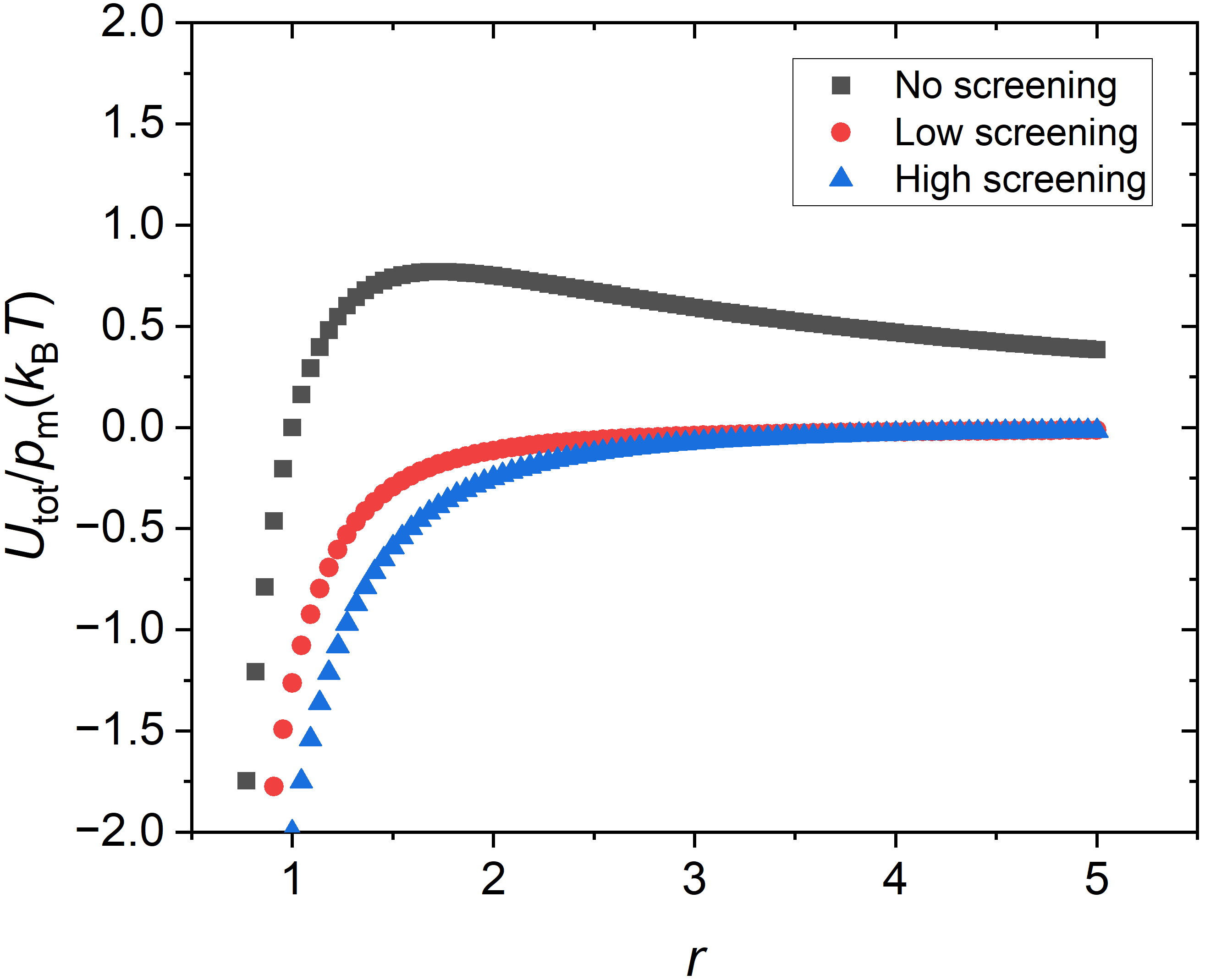}
\caption{Total interaction potential of a pair of charged magnetic discs, oriented head to tail in a simplified model of the effects of increased Debye screening. The $x$-axis represents the distance between disc centers, while the $y$-axis shows the potential energy, normalised by the magnetic dipole-dipole interaction prefactor. The three curves show that as the screening is increased, the potential goes from repulsive to mildly attractive.}
    \label{fgr:potentials}
\end{figure}

To illustrate the effects of altering the DBSA concentration on degree of self-assembly in the colloidal suspension, Figure \ref{fgr:potentials} shows how even a simple approximation of the interparticle interactions is strongly affected by changes in the screening length. The dipolar interaction between two magnetic nanoplatelets with magnetic moment $\boldsymbol{\mu}_i$ and separated by the vector $\mathbf{r}$ can be written as:

\begin{equation}
U_{\mathrm{dd}} (\mathbf{r}) = \frac{\mu_0}{4 \pi} \Big(  \frac{(\boldsymbol{\mu}_i \boldsymbol{\mu}_j)}{|r|^3} - \frac{3(\boldsymbol{\mu}_i \mathbf{r})(\boldsymbol{\mu}_j \mathbf{r})}{|r|^5} \Big)
\end{equation}

and the Coulomb potential with added Yukawa screening term can written as:

\begin{equation}
U_{\mathrm{C-DH}} (\mathbf{r}) = \frac{1}{4 \pi \epsilon_0}  \frac{q_i \cdot q_j}{|r|} \exp{(-\kappa |r|)}
\end{equation}
 
\noindent where $q_i$ and $q_j$ are the particle charges and $\kappa$ is the Debye screening constant. 
For this approximation, we assume that the two platelets are of the same size, and have the same magnetic moment and effective charge, which reduces the total potential to:

\begin{equation}
U_{\mathrm{tot}} (\mathbf{r}) = p_{\mathrm{m}}  \cdot \frac{-2 \mu^2 }{r^3} + p_{\mathrm{c}} \cdot \frac{q^2}{r} \exp{(-\kappa r)}
\end{equation}

\noindent where $p_{\mathrm{m}}$ and $p_{\mathrm{c}}$ are scaling factors which encompass the relative strength of the magnetic and electrostatic interactions. Based on experimental observations, we fix $p_{\mathrm{c}} > p_{\mathrm{m}}$ and vary $\kappa$ to approximate the effect of added DBSA. 

Although the simplifying assumptions in Figure \ref{fgr:potentials} are too strong to claim that this potential quantitatively represents the experimental system - both the high degree of polydispersity and the more complex electrostatics would need to be considered in greater detail - it qualitatively shows that gradually increasing the screening length will destabilise the system, leading to self-assembly.

To understand the influence of the ratio ${c_\mathrm{{MP}}/c_\mathrm{{DBSA}}}$  on the dynamic magnetic response, in this work, we explored magnetic dynamics in two series of suspensions: (i) a series with low but constant magnetic particle concentration ($c_\mathrm{{MP}}=\qty{8}{\gram\per\liter}$) and varying $c_\mathrm{{DBSA}}$ (Set \textbf{1}) and (ii) a dilution series where we kept the $c_\mathrm{{MP}}/c_\mathrm{{DBSA}}$ ratio constant (Set \textbf{2}). The latter was chosen so that the ferrofluid with the lowest platelet concentration %(8~mg/ml) 
($c_\mathrm{{MP}}=\qty{8}{\gram\per\liter}$) 
does not give rise to collective modes in the ACS spectra.
The charge and the ionic strength determine the electrostatic environment of the ferrofluid: the desorption of the surfactant molecules from the positively charged platelets and the dissociation of the dissolved DBSA. On the first approach, we presume that keeping the ${c_\mathrm{{MP}}}/{c_\mathrm{{DBSA}}}$ constant, the electrostatic environment does not change significantly (the adsorption-desorption equilibrium of the surfactant is not shifted) and the response would only depend on the magnetic particle concentration i.e the magnetostatic interactions, hence the study of Set \textbf{2}.

In the following we discuss collective modes of magnetic dynamics in Sc-BaHF suspensions. We show, that even at low magnetic particle concentration, as in Set \textbf{1} ($c_\mathrm{MP}=\qty{8}{\gram\per\liter}$), collective modes emerge on increasing surfactant concentration ($c_\mathrm{DBSA}$). Besides, investigating Set \textbf{2}, we also demonstrate the role of the magnetic particle concentration ($c_\mathrm{MP}$) in the collective behaviour.

\section{Materials and methods}
\label{sec:sample1}

\subsection{Ferrofluid preparation}
The stock ferrofluid was prepared as described in \cite{Kuester.2022}. The sets of ferrofluids were prepared by diluting the stock suspension ($c_\mathrm{MP}=\qty{304}{\gram\per\liter}$ and $c_\mathrm{DBSA}=\qty{43}{\gram\per\liter}$) to the necessary concentration of MPs and DBSA with appropriate DBSA solution in 1-butanol and/or by 1-butanol. The resulting suspensions were sonicated for a minute. Set \textbf{1} marks the suspensions with constant $c_\mathrm{MP}=\qty{8}{\gram\per\liter}$ and varying $c_\mathrm{DBSA}$. They are designated as $\mathrm{MP8}$ and the corresponding weight percent of DBSA as in Table \ref{tbl:materials}. Set \textbf{2} identifies the dilution series with varied $c_\mathrm{MP}$ but constant $c_\mathrm{MP}/c_\mathrm{DBSA}$. They are designated with the abbreviation MP and the corresponding magnetic particle concentration as in Table \ref{tbl:materials}.

\begin{table}[]
\centering
\begin{tabular}{llllll}
\hline
\multicolumn{3}{|c|}{\textbf{Set 1} $c_\mathrm{MP}$=$\qty{8}{\gram\per\liter}$}                                                                                           & \multicolumn{3}{c|}{\textbf{Set 2} $c_\mathrm{MP}/c_\mathrm{DBSA}$=6.7}                                                                                      \\ \hline
\multicolumn{1}{|c|}{Designation} & \multicolumn{1}{c|}{\%DBSA} & \multicolumn{1}{c|}{$\phi_\mathrm{MP}$/$\phi_\mathrm{DBSA}$} & \multicolumn{1}{c|}{Designation} & \multicolumn{1}{c|}{$c_\mathrm{MP}/\mathrm{gL^{-1}}$} & \multicolumn{1}{c|}{$\phi_\mathrm{MP}$} \\ \hline
\multicolumn{1}{|c|}{MP8\_12}     & \multicolumn{1}{c|}{12}     & \multicolumn{1}{c|}{1.44}                                    & \multicolumn{1}{c|}{\cellcolor{yellow}MP8}         & \multicolumn{1}{c|}{8}                      & \multicolumn{1}{c|}{0.0015}             \\ \hline
\multicolumn{1}{|c|}{\cellcolor{yellow}MP8\_13}     & \multicolumn{1}{c|}{13}     & \multicolumn{1}{c|}{1.37}                                    & \multicolumn{1}{c|}{MP12}        & \multicolumn{1}{c|}{12}                     & \multicolumn{1}{c|}{0.0024}             \\ \hline
\multicolumn{1}{|c|}{MP8\_16}     & \multicolumn{1}{c|}{16}     & \multicolumn{1}{c|}{1.04}                                    & \multicolumn{1}{c|}{MP32}        & \multicolumn{1}{c|}{32}                     & \multicolumn{1}{c|}{0.0063}             \\ \hline
\multicolumn{1}{|c|}{MP8\_21}     & \multicolumn{1}{c|}{21}     & \multicolumn{1}{c|}{0.73}                                    & \multicolumn{1}{c|}{MP40}        & \multicolumn{1}{c|}{40}                     & \multicolumn{1}{c|}{0.0077}             \\ \hline
\multicolumn{1}{|c|}{MP8\_25}     & \multicolumn{1}{c|}{25}     & \multicolumn{1}{c|}{0.61}                                    & \multicolumn{1}{c|}{MP92}        & \multicolumn{1}{c|}{92}                     & \multicolumn{1}{c|}{0.0170}             \\ \hline
\multicolumn{1}{|c|}{MP8\_26}     & \multicolumn{1}{c|}{26}     & \multicolumn{1}{c|}{0.57}                                    & \multicolumn{1}{c|}{MP126}       & \multicolumn{1}{c|}{126}                    & \multicolumn{1}{c|}{0.0229}             \\ \hline
\multicolumn{1}{|c|}{MP8\_50}     & \multicolumn{1}{c|}{50}     & \multicolumn{1}{c|}{0.20}                                    & \multicolumn{1}{c|}{MP158}       & \multicolumn{1}{c|}{158}                    & \multicolumn{1}{c|}{0.0285}             \\ \hline
                                     
\end{tabular}
\caption{In Set \textbf{1} the magnetic particle concentration $c_\mathrm{MP}=m_\mathrm{MP}/V_\mathrm{1-BuOH}$ is fixed to $\qty{8}{\gram\per\liter}$. The DBSA concentration is given as $\%\mathrm{DBSA}=m_\mathrm{DBSA}/(m_\mathrm{MP}+m_\mathrm{DBSA})$.
The volume percents apply to the total volume of the ferrofluid.
In the dilution series Set \textbf{2} the concentration ratio $c_\mathrm{MP}/c_\mathrm{DBSA}$ is fixed to 6.7. The concentration $c_\mathrm{MP}=m_\mathrm{MP}/V_\mathrm{1-BuOH}$ is utilized to designate the suspensions. The volume fraction $\phi_\mathrm{MP}=V_\mathrm{MP}/V_\mathrm{FF}$. MP8\_13 of Set \textbf{1} and MP8 of Set \textbf{2} are identical suspensions.}
\label{tbl:materials}
\end{table}

\subsection{AC susceptometry}
Measurements of the AC susceptibility were carried out with a custom-made setup, which was originally built for measurements of the dynamics of magnetic nanoparticles in a rotating magnetic field \cite{Dieckhoff:2011da}. For the ACS measurements just one set of Helmholtz coils was used to generate the sinusoidal excitation field. ACS spectra were recorded at \qty{298}{\kelvin} in a frequency range between \qty{0.1}{\hertz} and \qty{2.2}{\kilo\hertz} and at field amplitudes between \qty{0.5}{\milli\tesla} and \qty{5}{\milli\tesla}. 

The exposure of a ferrofluid to an alternating magnetic field results in aligned magnetisation, which will oscillate with a phase lag. The dynamic response can be described by the real $\chi'$ and imaginary $\chi''$ parts of the magnetic susceptibility. The ACS susceptibility spectra of MP suspensions are generally analysed with the Debye model where the complex susceptibility is given by

\begin{equation}
\chi (\omega)=\frac{\chi_0}{1+i\omega\tau}
\end{equation}
with the static susceptibility $\chi_{0}$, the angular frequency $\omega=2\pi f$ and the characteristic relaxation time of the MP $\tau$. Since the Sc-BaHF platelets are thermally blocked \cite{Kuester.2022}, only Brownian relaxation occurs. The Brownian relaxation time \cite{Valiev:2007hn} $\tau_{\mathrm{B}}$ is determined by the viscosity of the medium $\eta$, the hydrodynamic volume of the MP $V_{\mathrm{H}}$ and the temperature $T$:
\begin{equation}
    \tau_{\mathrm{B}}=\frac{3\eta V_{\mathrm{H}}}{k_{\mathrm{B}}T}
\label{eq:Brownian}
\end{equation}

The field dependence of the Brownian relaxation time is well described by the empirical model by Yoshida and Enpuku \cite{Yoshida:2009jc}
in the frame of the Fokker-Planck formalism. For non-interacting thermally blocked MPs, their model is valid for magnetic field amplitudes with a Langevin parameter $\xi=m \mu_0 H /\left(k_{\mathrm{B}} T\right)$ of up to 300.
The relaxation time is given by: 

\begin{equation}
\tau_{\mathrm{B,H}}=\frac{\tau_{\rm{B,0}}}{\sqrt{1+0.126 \xi^{1.72}}},
\label{eq:ye}
\end{equation} 
where $\tau_{\mathrm{B,0}}$ is the relaxation time in the limit $H\rightarrow 0$.

To account for a distribution of relaxation times due to the polydispersity of MPs and interparticle interactions, we apply the Cole-Cole equation \cite{Cole}.

\begin{equation}
    \chi(f)=\chi_{\infty}+\frac{\Delta \chi}{1+\left(i \omega \tau_{\mathrm{B,H}}\right)^{1-\alpha}}
\label{eq:CC1}
\end{equation}
considering symmetric broadening only. Here $\chi_{\infty}$ is the susceptibility in the high-frequency limit, $\Delta\chi$ is proportional to the amplitude of the susceptibility, and $\alpha$ is the broadening coefficient ($\alpha=0$ corresponds to the Debye model). In order to analyse ACS spectra with multiple relaxation modes, the experimental data were fitted with a sum of several Cole-Cole equations:

\begin{equation}
    \chi(f)=\chi_{\infty}+\sum_i\frac{\Delta \chi_i}{1+\left(i \omega \tau_{\mathrm{B,H},i}\right)^{1-\alpha_i}}
\label{eq:CC}
\end{equation}

\noindent In our study, up to three modes $i$ could be resolved.

To match these experiment to further theoretical work, we will later fit the calculated Langevin susceptibility $\chi_L$ to an adaption of the susceptibility which includes chain formation~\cite{Mendelev:2004ve}. However, this model only was dervied for dipolar hard spheres, the shape of which (and thus the self-assembly process) significantly differs from that of spheres. Therefore, we have calculated a new expression for the partition function, the calculation of which is described in the SI). The resulting expression was then fitted using a least-squares fit to determine $\lambda$.

\section{Results}
\label{sec:The ACS spectra}
\subsection{Analysis of ACS spectra of Set \textbf{1}}
\begin{figure}[H]
\centering
  \includegraphics[width=0.8\columnwidth ]{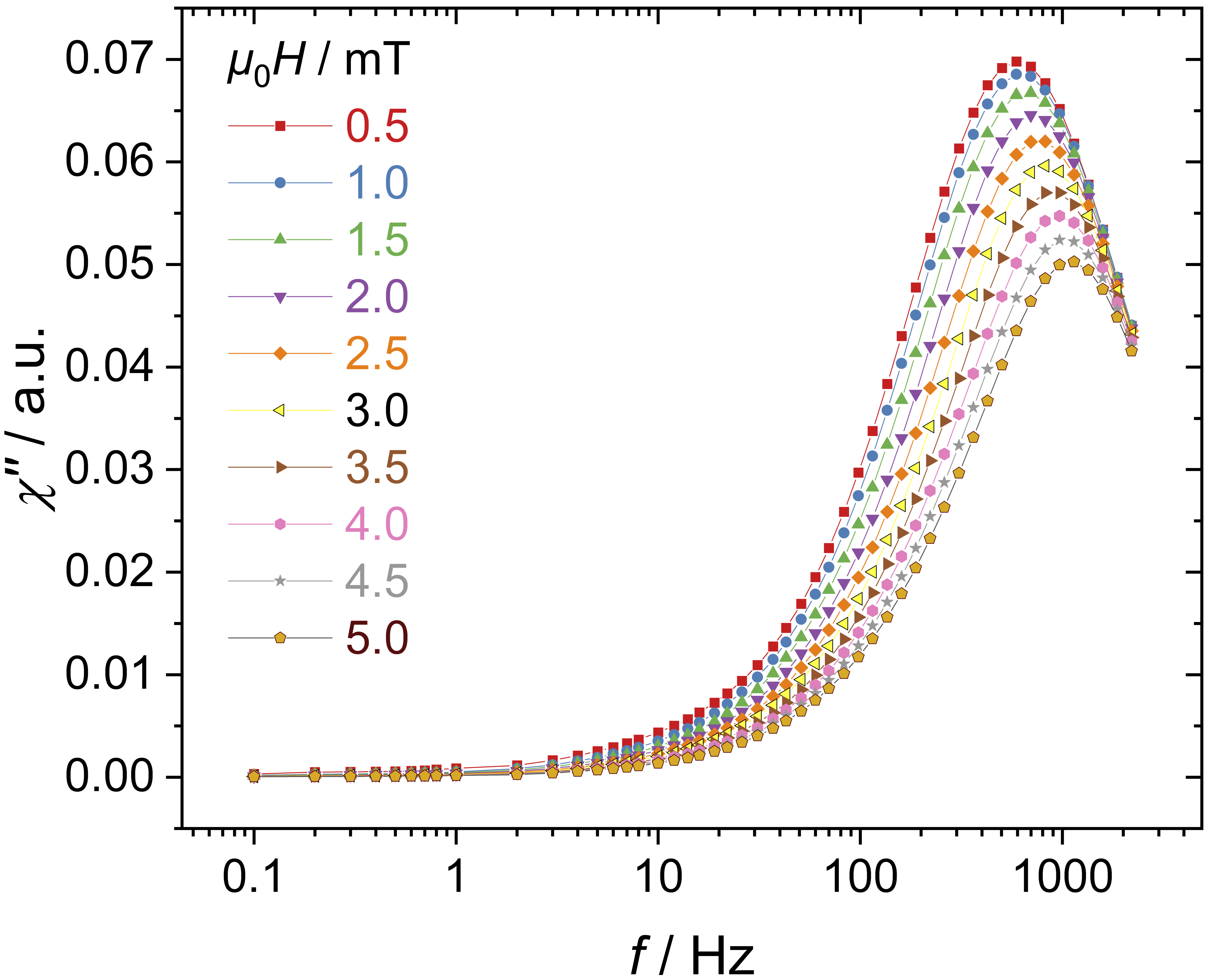}
  \caption{The ACS spectra of $\mathrm{MP}8\_13$ recorded for different amplitudes of the probe field consist of a single symmetric high-frequency peak related to the single-platelet relaxation mode. The maximum shifts to higher frequencies with increasing probe field.}
    \label{MP8_13_single peak}
\end{figure}

To explore the influence of the surfactant concentration on the magnetic response, we measured the set of ferrofluids  \textbf{1} containing a fixed low concentration of MPs ($\qty{8}{\gram\per\liter}$) and varied $c_{\mathrm{DBSA}}$. The ACS spectra of the suspensions where $\phi_\mathrm{MP}/\phi_\mathrm{DBSA}>1$ are characterised by a single  peak in the high-frequency range attributed to the relaxation of single platelets (Fig.~\ref{MP8_13_single peak}). In the following, we designate this relaxation mode as a high-frequency (HF) mode.
With increasing field amplitude the peaks' maximum shifts to higher frequencies and its amplitude decreases. Similar behaviour was observed in aqueous suspensions of spherical \cite{Dieckhoff:2016ev} and rod-like nanoparticles \cite{Remmer:2017gf}.

\begin{figure}[hbt!]
\centering
\subfloat[]{\includegraphics[width = 0.3\columnwidth]{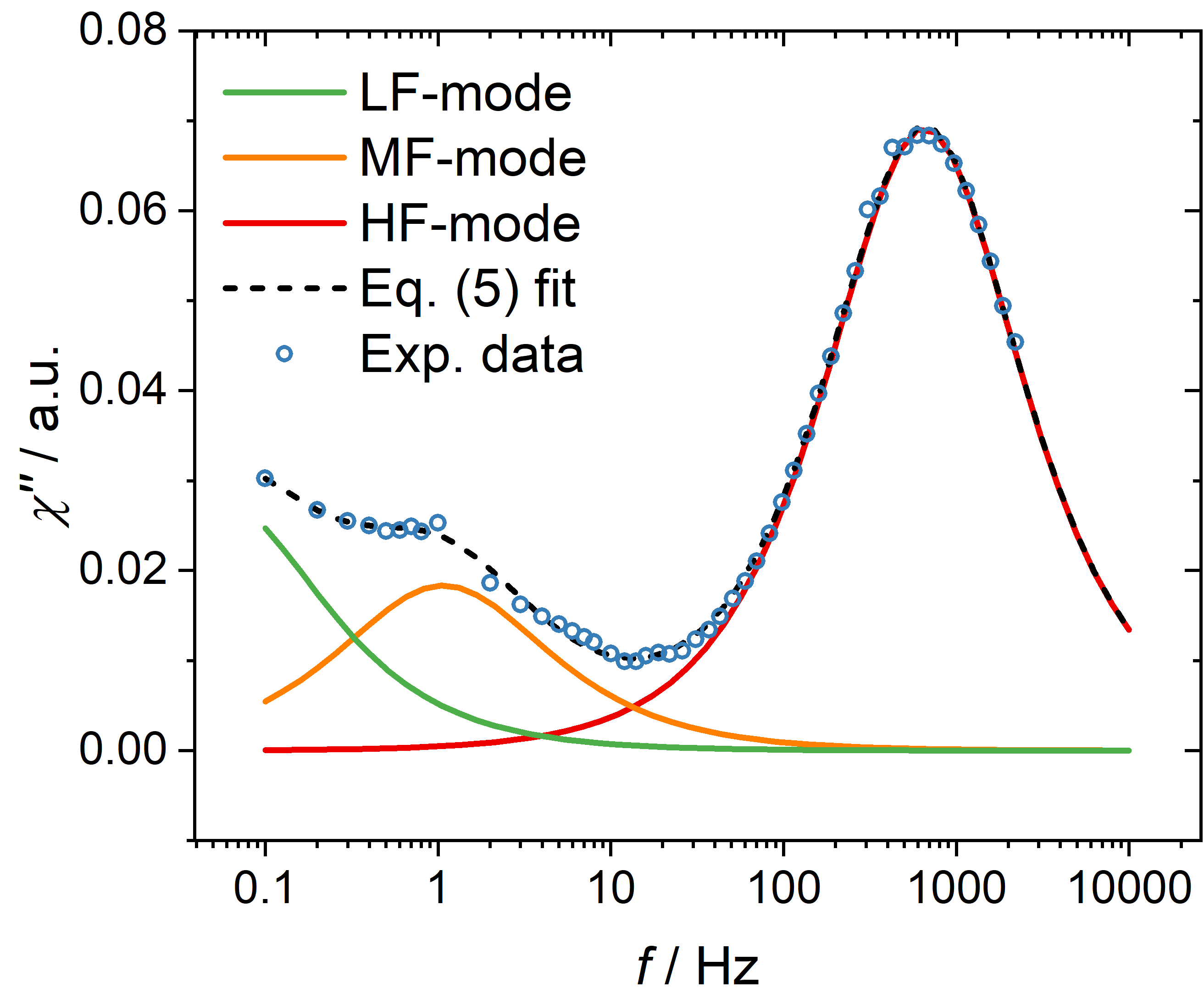}}
 \hspace{1em}
\subfloat[]{\includegraphics[width = 0.3\columnwidth]{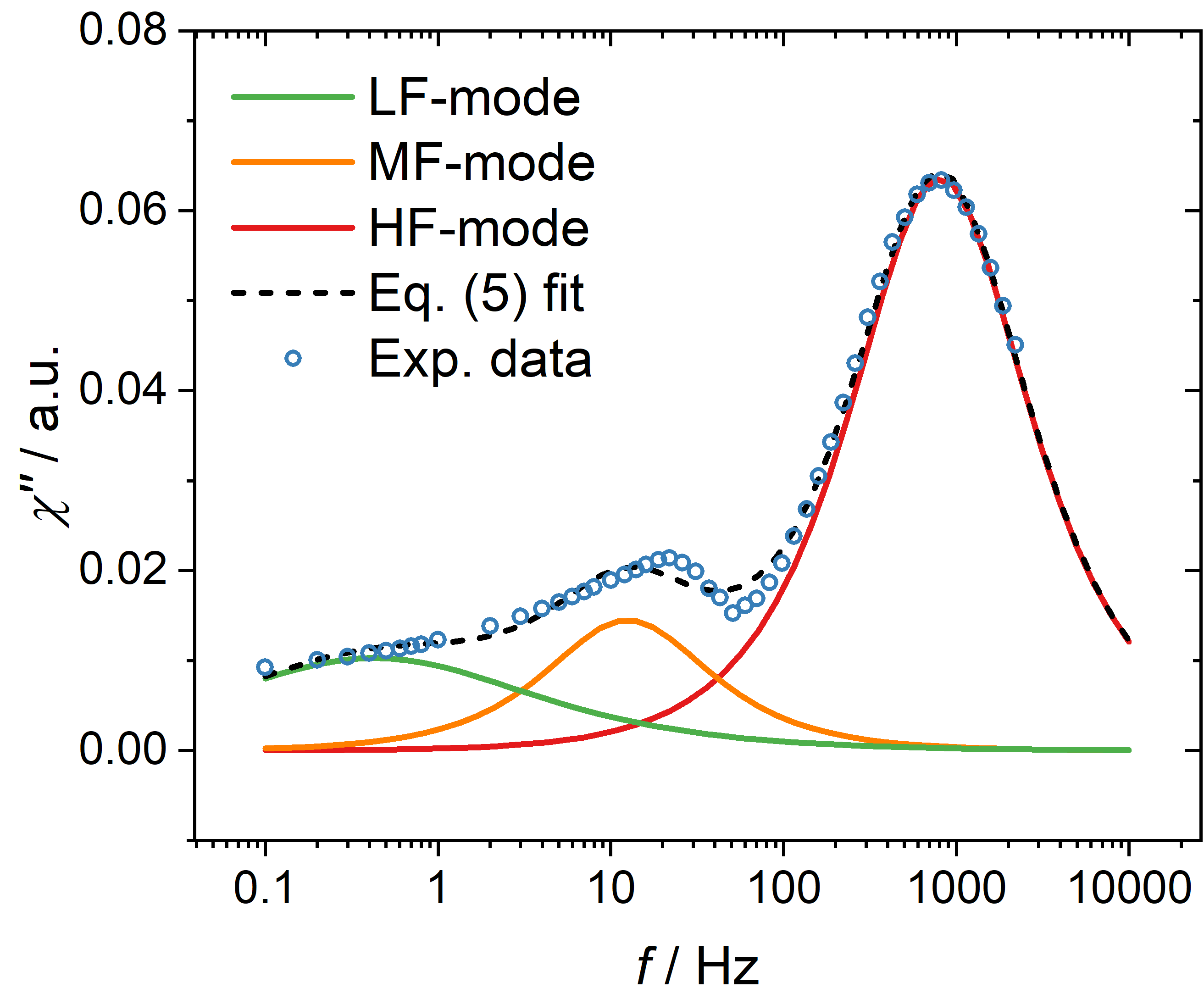}}
 \hspace{1em}
\subfloat[]{\includegraphics[width = 0.3\columnwidth]{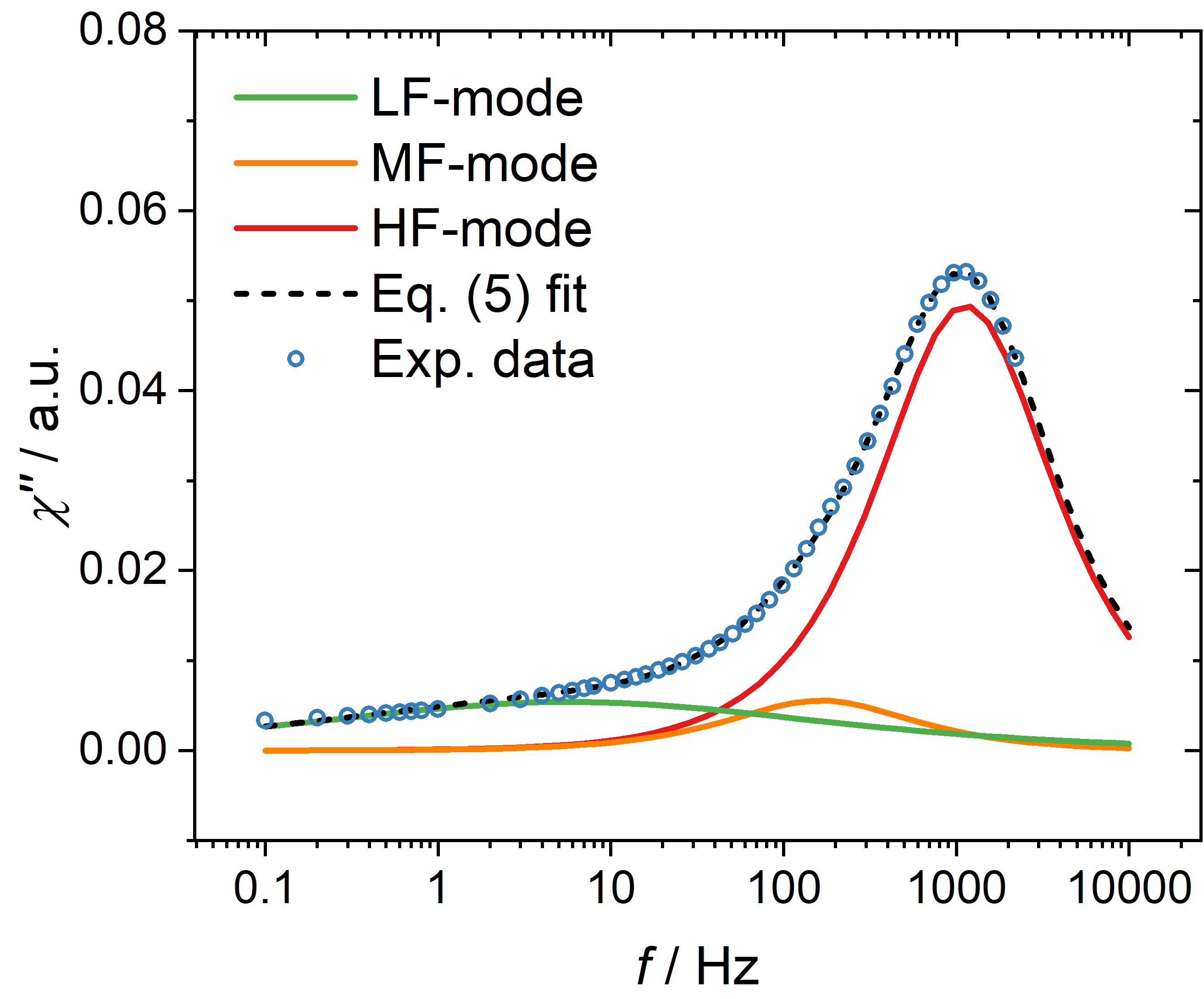}}
 
\caption{Multiple peaks of $\mathrm{MP}8\_26$ at (a) \qty{0.5}{\milli\tesla}, (b) \qty{2.5}{\milli\tesla} and (c) \qty{5}{\milli\tesla}. With increasing probe field, the middle- and high-frequency peaks collapse as a result of progressing mode-overlapping. The slowest mode (LF mode)
shifts to higher frequencies and broadens, transforming into a diminishing flattened peak.}
\label{MP8_26_ACS}
\end{figure}

A further increase of $c_{\mathrm{DBSA}}$ results in the emergence of low-frequency modes (Fig.~\ref{MP8_26_ACS}). The onset of the slowest mode (LF) can only be distinguished at low probe fields. The spectral maximum of this mode shifts to higher frequencies with increasing probe field while its amplitude decreases and the peak apparently flattens. This contribution acts effectively as an offset of the spectrum. 

An additional peak in the intermediate range, denoted as middle-frequency mode (MF), emerges in this concentration range. This mode merges into the HF mode with increasing probe field while the amplitude seemingly decreases. However, since the modes overlap, it is difficult to estimate the relaxation times accurately.

\begin{figure}[hbt!]
  \includegraphics[width=0.8\columnwidth]{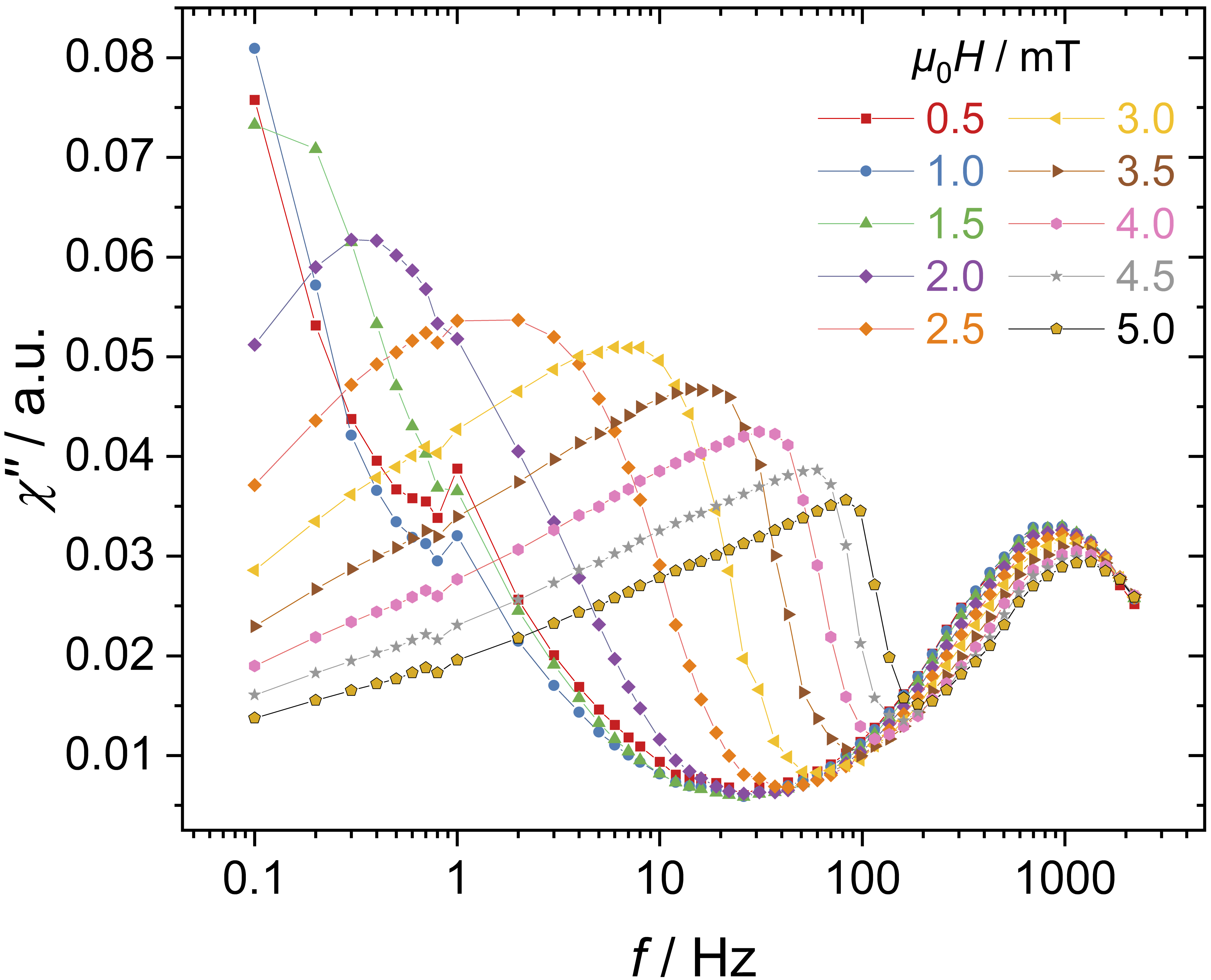}
  \caption{The ACS spectra of $\mathrm{MP}8\_50$ recorded for different amplitudes of the probe field are dominated by collective modes. The nearly linear in log-scale slope in the frequency range below \qty{100}{\hertz}} for AC fields between \qtyrange[range-phrase=~-~]{3}{5}{\milli\tesla} implies the complex spectral structure of multiple overlapping peaks.
    \label{fgr:MP8_50_ACSspectra}
\end{figure}

\begin{figure}[hbt!]
\subfloat[]{\includegraphics[width = 0.50\columnwidth]{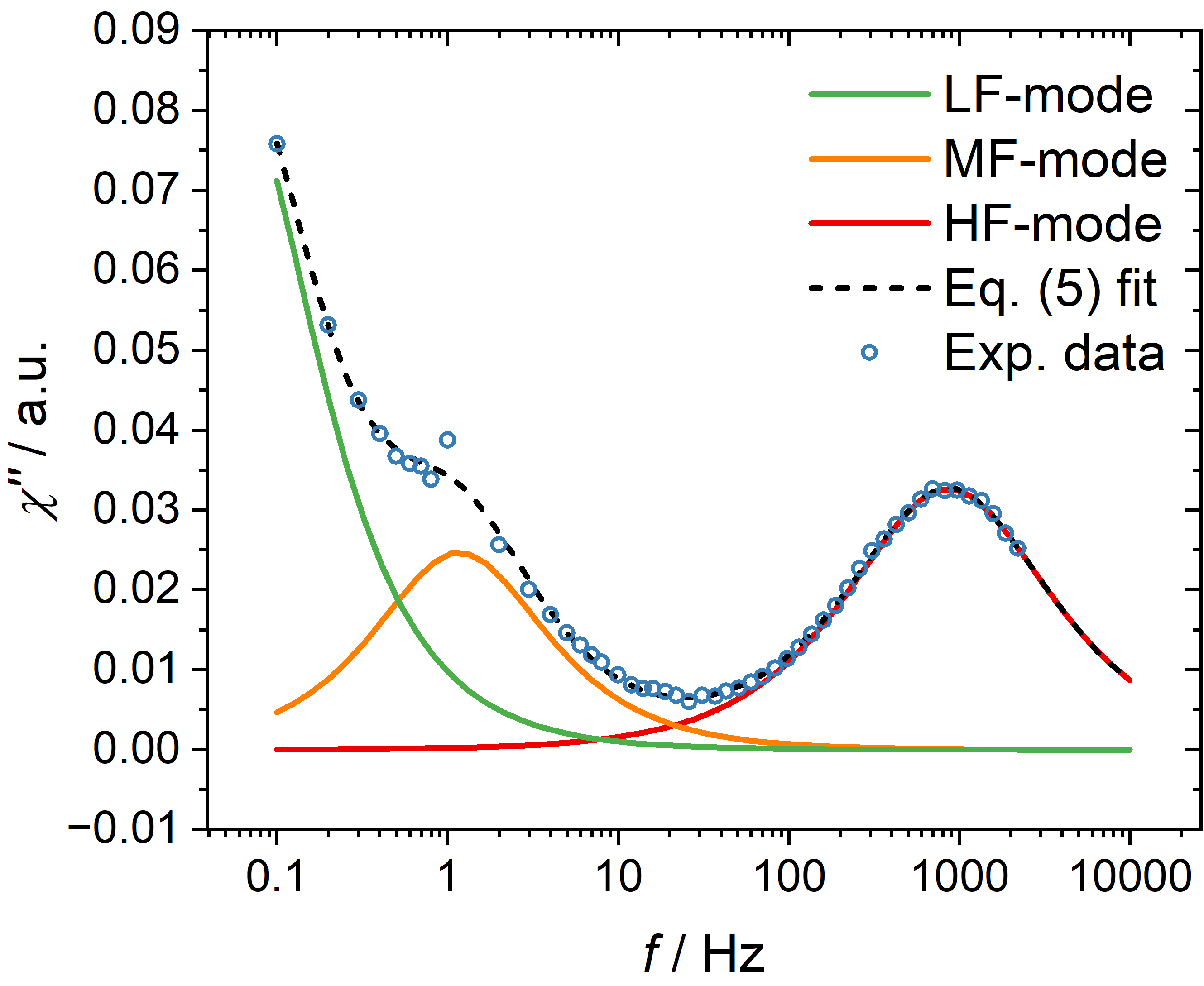}} 
 \hspace{2em}
\subfloat[]{\includegraphics[width = 0.50\columnwidth]{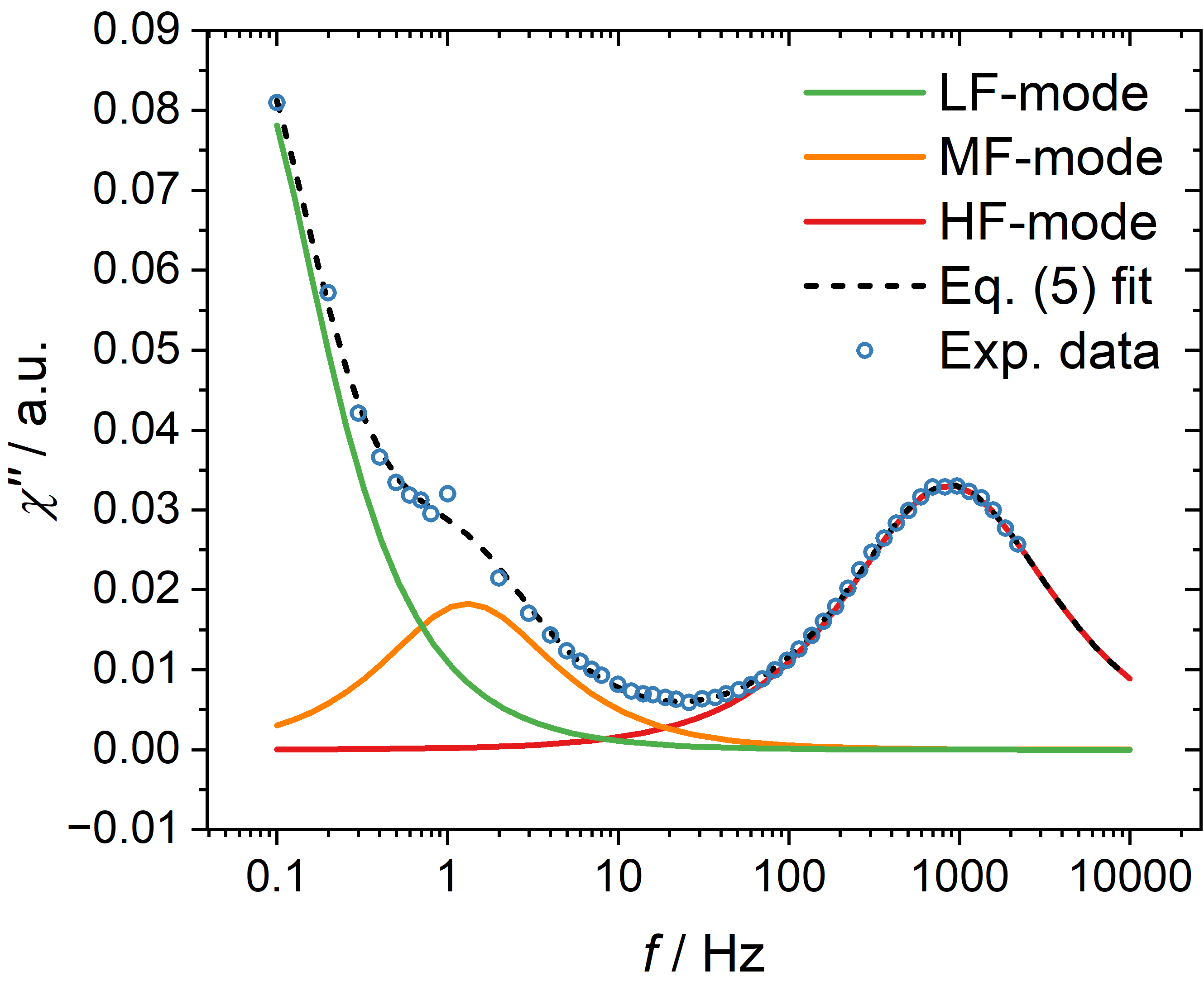}}
\caption{Multiple peaks in spectra of $\mathrm{MP}8\_50$ recorded in probe fields with amplitude \qty{0.5}{\milli\tesla} (a), and  and \qty{1}{\milli\tesla} (b) with dominating but well distinguishable low- and middle-frequency modes.}
\label{MP8_50_0p5_1mT_fit}
\end{figure}

The ACS spectra of $\mathrm{MP8\_50}$ have very distinctive features (Fig.~\ref{fgr:MP8_50_ACSspectra}). 
They consist of a very pronounced HF peak and especially prominent low-frequency collective modes. Already in \qty{0.5}{\milli\tesla} probe field, the remaining LF peak is significantly stronger than the HF peak maximum. In \qty{0.5}{\milli\tesla} and \qty{1}{\milli\tesla} fields, one can fit three peaks that can be assigned to the LF, MF and HF modes (Fig.~\ref{MP8_50_0p5_1mT_fit}). In higher fields, we observe a nearly linear increase of the imaginary part of the magnetic susceptibility as a function of $\log f$ and an HF peak. Its mode structure could only be implicitly determined using Eq.~(\ref{eq:CC}) over a reduced frequency interval. Multiple overlapping in the ACS spectra of $\mathrm{MP8\_50}$ in high probe fields does not allow accurate Cole-Cole fits. As a result, we cannot accurately determine the position of the HF peak and its dependence on the probe field amplitude from the fits.

The ACS measurement results described above are in qualitative agreement with the findings in \cite{Bostjancic.2019}. At a sufficiently low DBSA concentration (12-16\%; cf. Table \ref{tbl:materials}), we observe just an HF mode in the ACS spectra, which we attribute to the Brownian rotation of single platelets, while at higher DBSA concentrations, collective modes appear, which indicate that the electrostatic repulsions are less effective to prevent the platelets from clustering.

\subsubsection{Field dependence of the Brownian relaxation time of the HF mode}
\label{The Yoshida-Enpuku fits}

\begin{figure}[hbt!]
\centering
\subfloat[]{\includegraphics[width = 0.3\columnwidth]{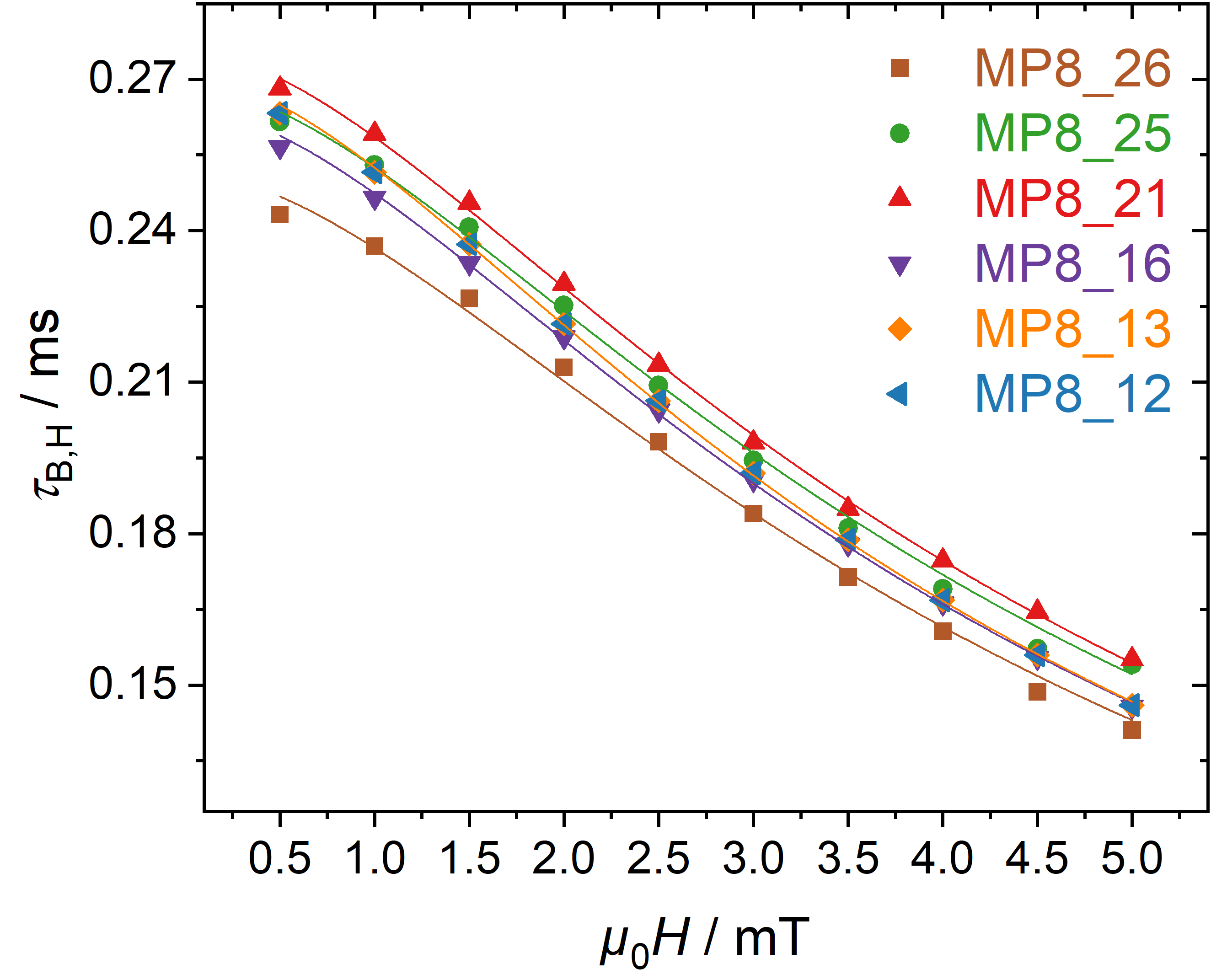}}
 \hspace{1em}
\subfloat[]{\includegraphics[width = 0.3\columnwidth]{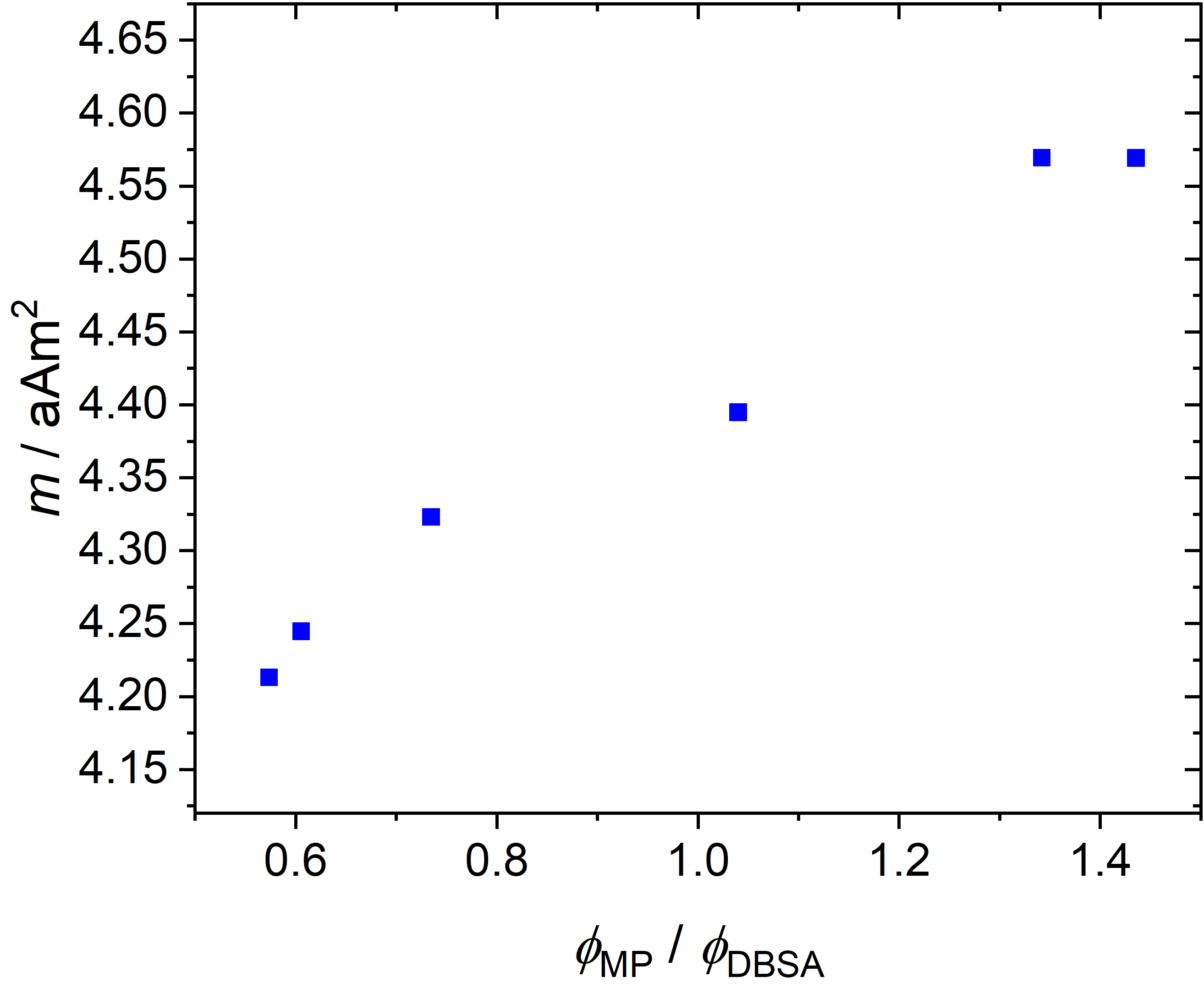}}
 \hspace{1em}
\subfloat[]{\includegraphics[width = 0.3\columnwidth]{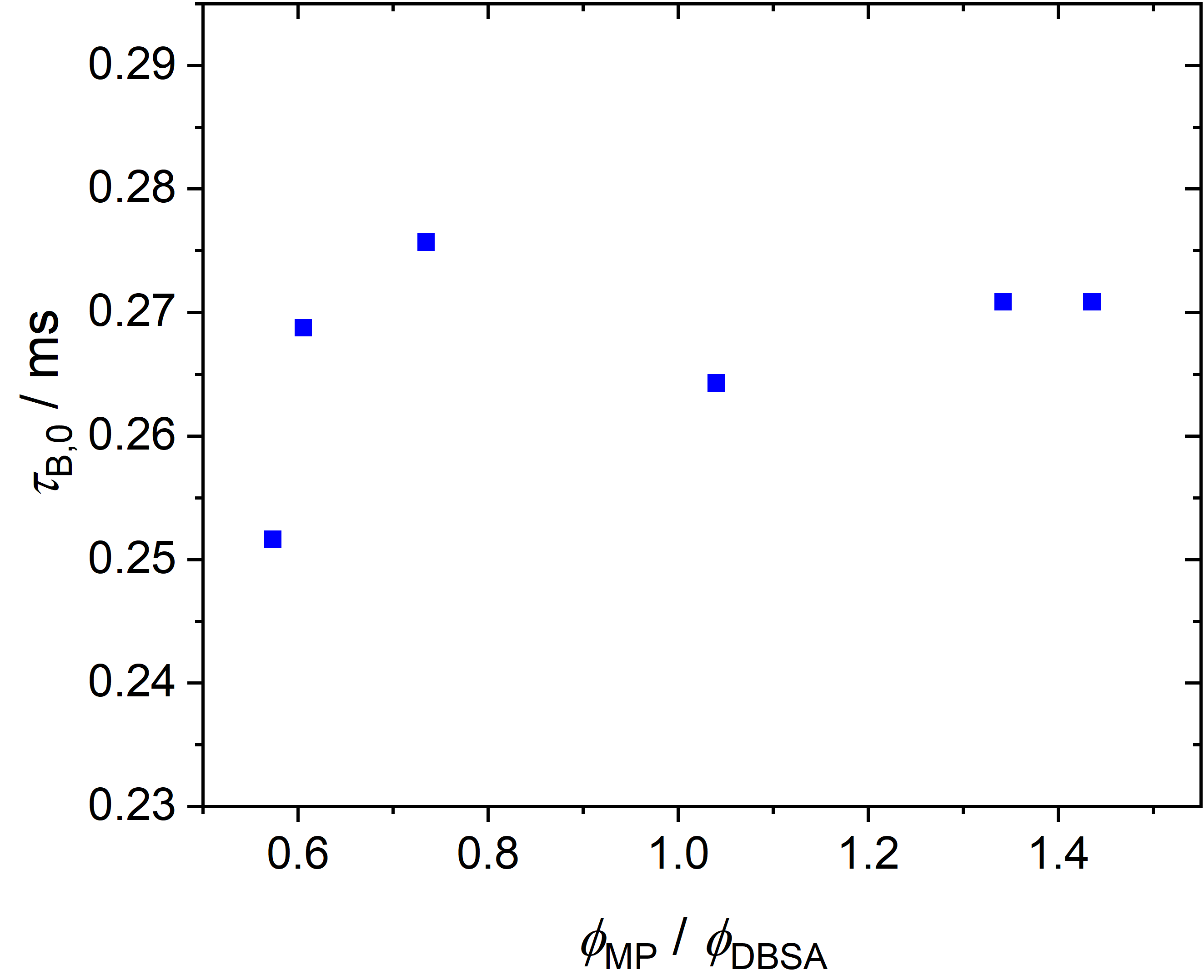}}

\caption{(a) Fit of the field dependence of the relaxation time of the HF mode using the model by Yoshida and Enpuku (Eq.~(\ref{eq:ye})) for Set \textbf{1}. The extracted magnetic moment
(b) and field-free relaxation time values (c) as a function of the concentration ratio of the magnetic particles to the surfactant DBSA.}
\label{YE_MP8}
\end{figure}

The Brownian relaxation times $\tau_{\mathrm{B,H}}$ estimated from the position of the HF peak in the ACS spectra as a function of the applied field amplitude are depicted in Fig.~\ref{YE_MP8}(a) along with the fits with  Eq.~(\ref{eq:ye}).
Since $\tau_{\mathrm{B,0}}$ and $\xi$ are treated as free parameters, the fitting allows one to determine the field-free Brownian relaxation time $\tau_{\mathrm{B,0}}$ and the magnetic moment $m$ for the individual samples. The results are depicted in Figs.~\ref{YE_MP8}(b, c). 

As shown in Fig.~\ref{YE_MP8}(b), the effective magnetic coupling as measured via $m$ slightly decreases with increasing amount of DBSA. Since the magnetic moment of the particles directly corresponds to their size, the gradual decrease indicates that with increasing surfactant concentration means the successively smaller particles are responding to the field. As Figs.~\ref{MP8_26_ACS}-\ref{MP8_50_0p5_1mT_fit} show, additional low-frequency relaxation modes appear at higher DBSA concentrations, which can be attributed to MP clusters. 

Due to the polydispersity of the platelets, these observations suggest that smaller platelets can still freely rotate contributing to the HF peak, while larger platelets form clusters. The remaining platelets contributing to the high-frequency range of the spectrum are, on average, smaller and thus have smaller magnetic moments $m$, resulting in shorter relaxation times. 

\subsection{Analysis of ACS spectra of Set \textbf{2}}
\begin{figure}[hbt!]
\centering
\subfloat[]{\includegraphics[width = 0.45\columnwidth]{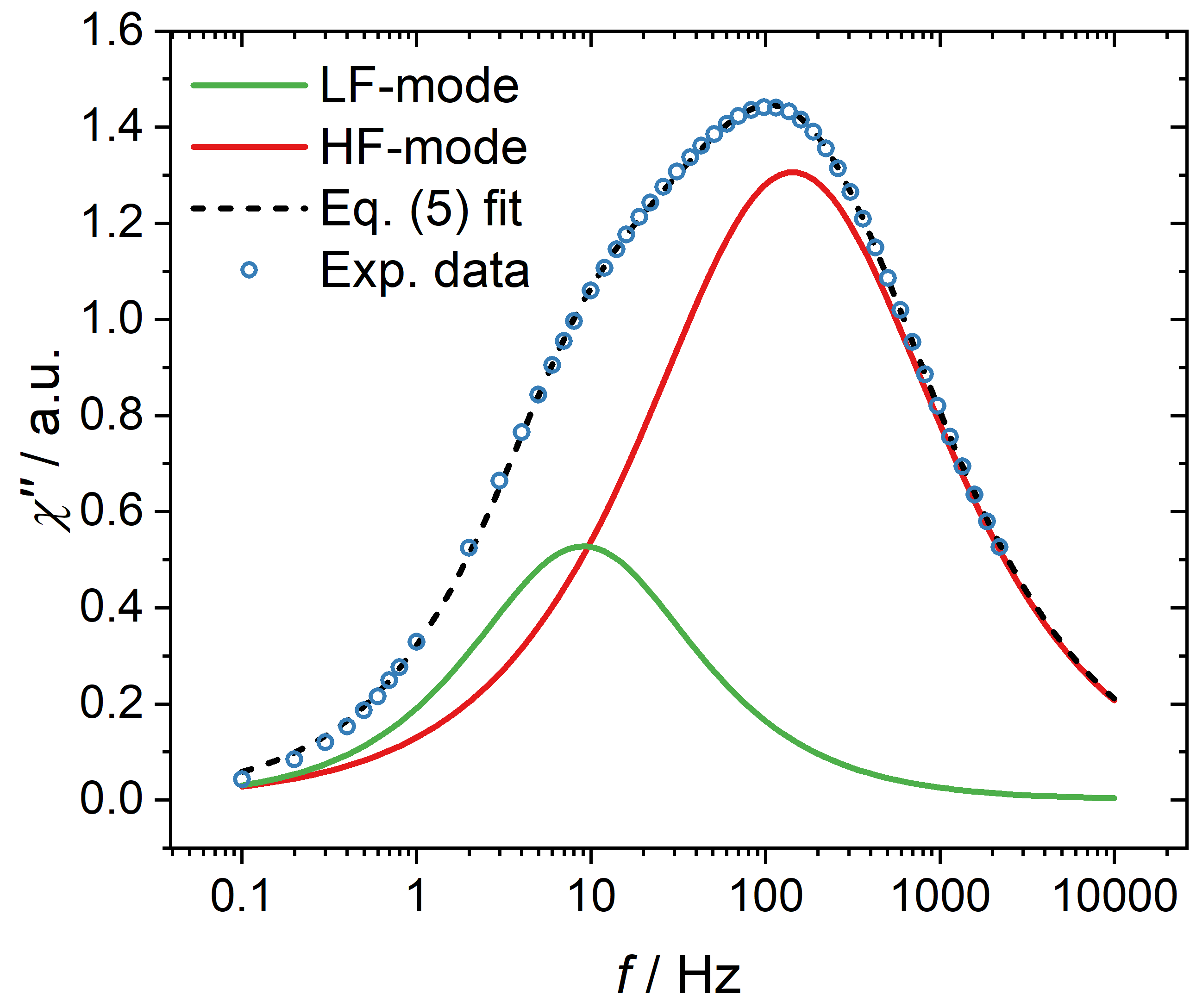}}
 \hspace{1em}
\subfloat[]{\includegraphics[width = 0.45\columnwidth]{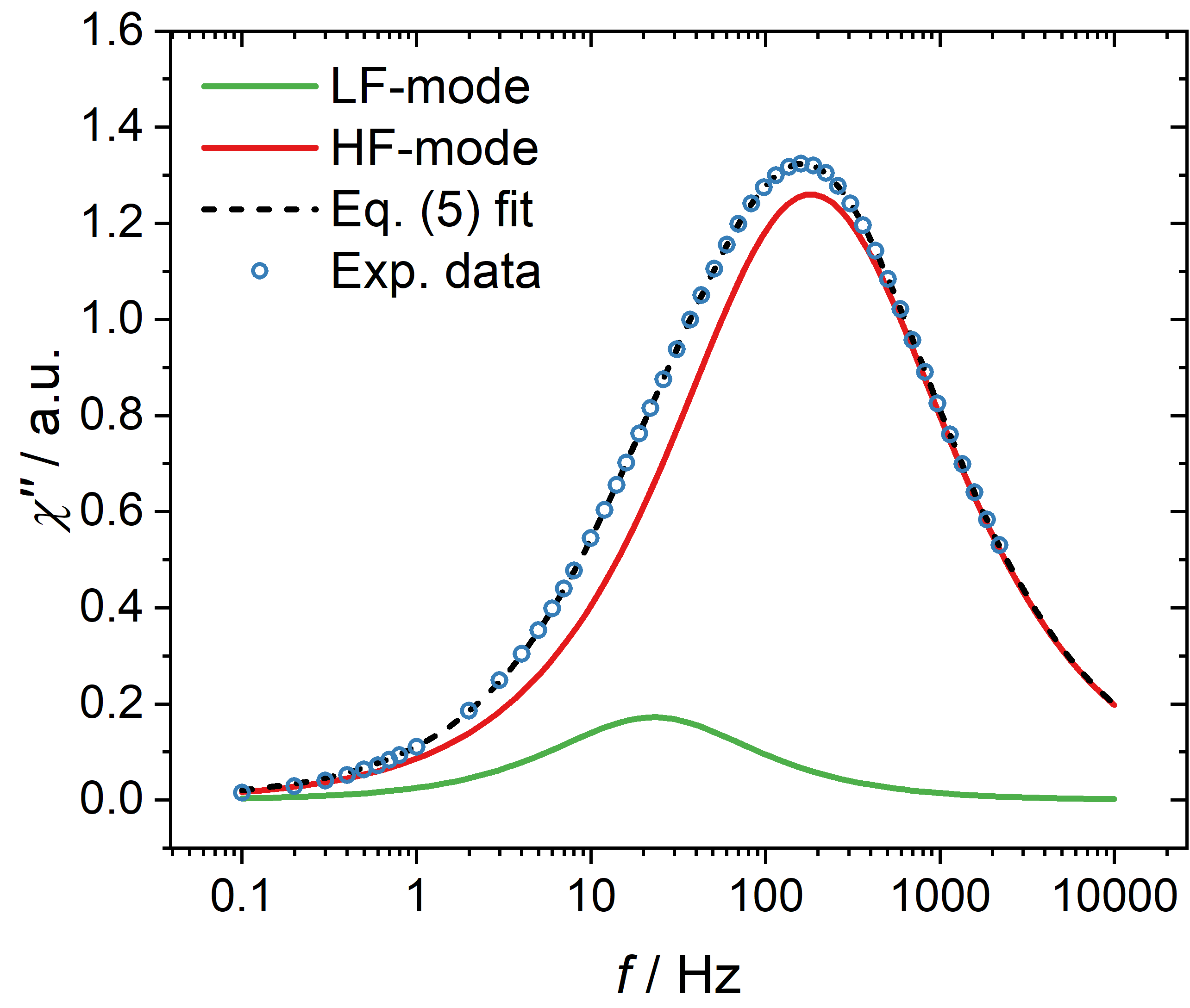}}\\
\subfloat[]{\includegraphics[width = 0.45\columnwidth]{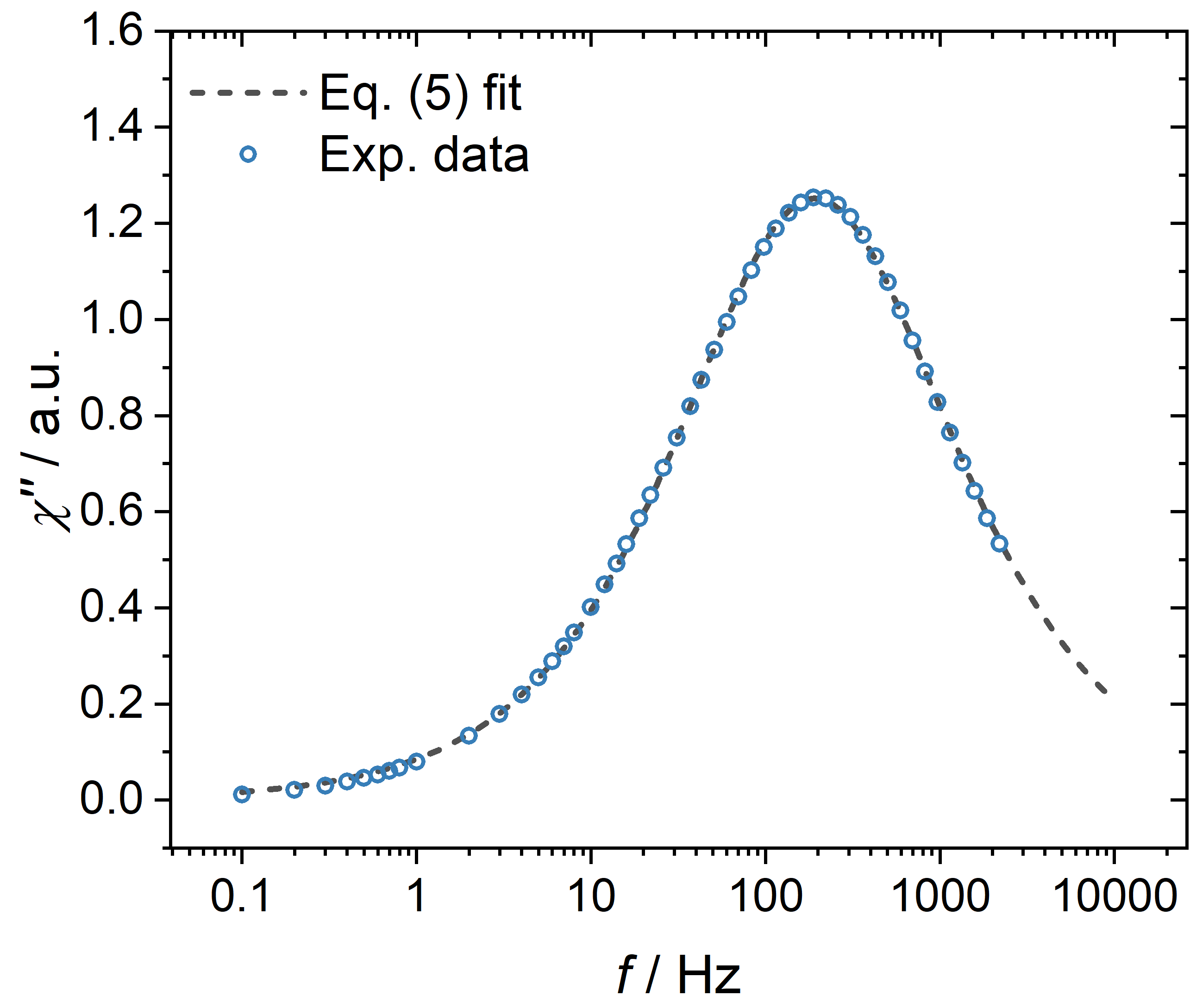}}
 \hspace{1em}
\subfloat[]{\includegraphics[width = 0.45\columnwidth]{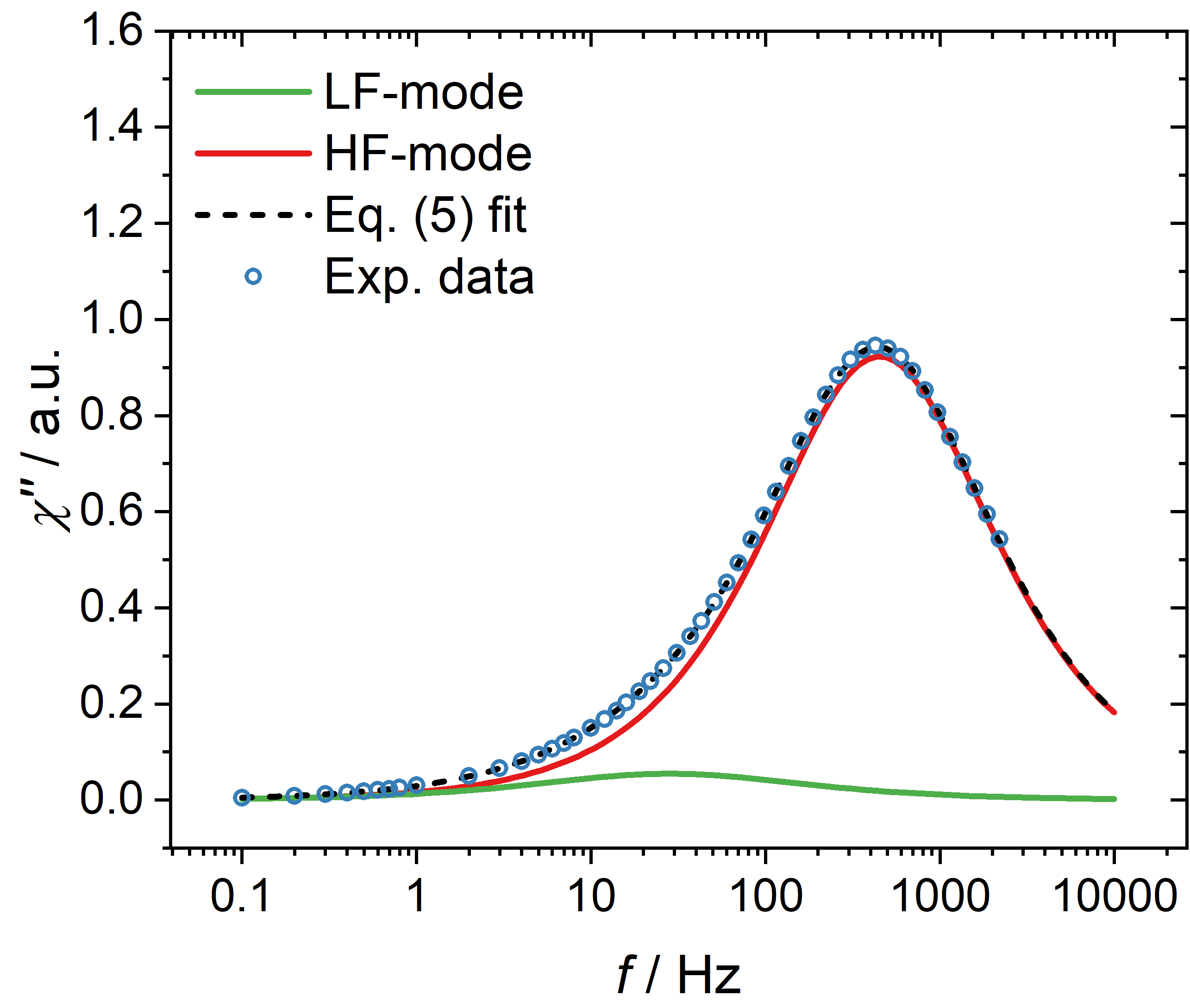}} 
\caption{Cole-Cole fits of the ACS spectra of $\mathrm{MP126}$ in AC fields of (a) \qty{0.5}{\milli\tesla}, (b) \qty{1.5}{\milli\tesla}, (c) \qty{2}{\milli\tesla} and (d) \qty{4.5}{\milli\tesla}. The low-frequency collective mode shifts with increasing field amplitude much stronger to higher frequencies than the HF peak does, so that at \qty{2}{\milli\tesla} the ACS spectrum consists of one symmetric peak. At an even higher field amplitude of \qty{4.5}{\milli\tesla}, the nematic mode emerges as a new low-frequency mode.}
\label{MP126_Cole-Cole}
\end{figure}

As the results of Set \textbf{1} indicate, a ratio ${c_\mathrm{{MP}}/c_\mathrm{{DBSA}}}=6.7$ prevents the Sc-BaHF platelets - at least at comparably low $c_\mathrm{MP}$ - from clustering.
The second set of magnetic fluids (Set \textbf{2}) was prepared on the simplified assumption that keeping the ratio of magnetic particles to the surfactant constant at 6.7 results in an unaltered electrostatic environment in the ferrofluids.
The spectra of the ferrofluids with the lowest MP concentrations, $\phi_\mathrm{{MP}}= 0.0015$ and 0.0024 (MP8 and MP12) have a single high-frequency peak corresponding to the single platelet relaxation (Fig.~\ref{MP8_13_single peak}). Upon further increasing the MP concentration to $\phi_\mathrm{{MP}}= 0.0063$ and 0.0077 (MP32 and MP40), a weak LF peak emerges at \qty{0.5}{\milli\tesla}, which is suppressed in stronger fields. First, at $\phi_\mathrm{{MP}}=0.0170$ (MP92), we could detect three modes in low probe fields (\qtyrange[range-phrase=~-~]{0.5}{1.5}{\milli\tesla}) and two modes in higher fields (\qtyrange[range-phrase=~-~]{2}{5}{\milli\tesla}). The LF mode again shows a very broad spectrum, as described above.
At higher concentration $\phi_\mathrm{{MP}}= 0.0229$ (MP126), the fits reveal two modes (Fig.~\ref{MP126_Cole-Cole}). In low probe fields (\qtyrange[range-phrase=~-~]{0.5}{1.5}{\milli\tesla}), there is an HF and an LF mode. With further increasing the probe field, the LF mode is not detectable anymore; the remaining peak is symmetric in shape.

However, in even higher fields (\qtyrange[range-phrase=~-~]{4.5}{5}{\milli\tesla}), the peak again becomes slightly asymmetric, and a new collective mode emerges. It has, nonetheless much lower amplitude than the low-frequency mode in low magnetic fields.

At the highest MP concentration $\phi_\mathrm{{MP}}= 0.0285$ (MP158), the character of the ACS spectra is similar, apart from the fact that the high probe field collective mode occurs already at \qty{4}{\milli\tesla}.
The effect of dipolar interactions on the ACS spectra at moderate MP concentrations has theoretically been studied by Ivanov and Camp. With increasing MP concentration, which affects the amount of self-assembly in the system and considering self-assembled clusters of MP, the position of the maximum in the ACS imaginary part shifts to smaller frequencies \cite{Camp:18}. Extending the work to higher dipolar coupling constants $\lambda \geq 4$, but still low MP concentrations, the authors found the appearance of additional peaks in the ACS imaginary part, which they attribute to the response of chains and rings \cite{Camp.2021}. While at low dipolar interaction parameters $\lambda$, the Brownian rotation of single MP dominate, at intermediate $\lambda$, the formation of particle chains and rings occurs with a peak frequency in the ACS imaginary part well below that of single MP. This seems to be the same underlying phenomenon as we find in our study, as dipolar magnetic nanoplatelets also form chains for at high values of $\lambda$ (or low electrostatic repulsion).
However, our study does not show a high-frequency peak (at about $24\omega\tau_{B}$), which they attribute to the motion of particles inside chains or rings. We exclude the latter mode for our case since no indication of such a mode was found in the measured ACS spectra- which makes sense given that platelets do not form rings and are sterically precluded from rotation inside chains.

\subsubsection{Field dependence of the Brownian relaxation time of HF mode}

\begin{figure}[hbt!]
\centering
\subfloat[]{\includegraphics[height = 5 cm]
{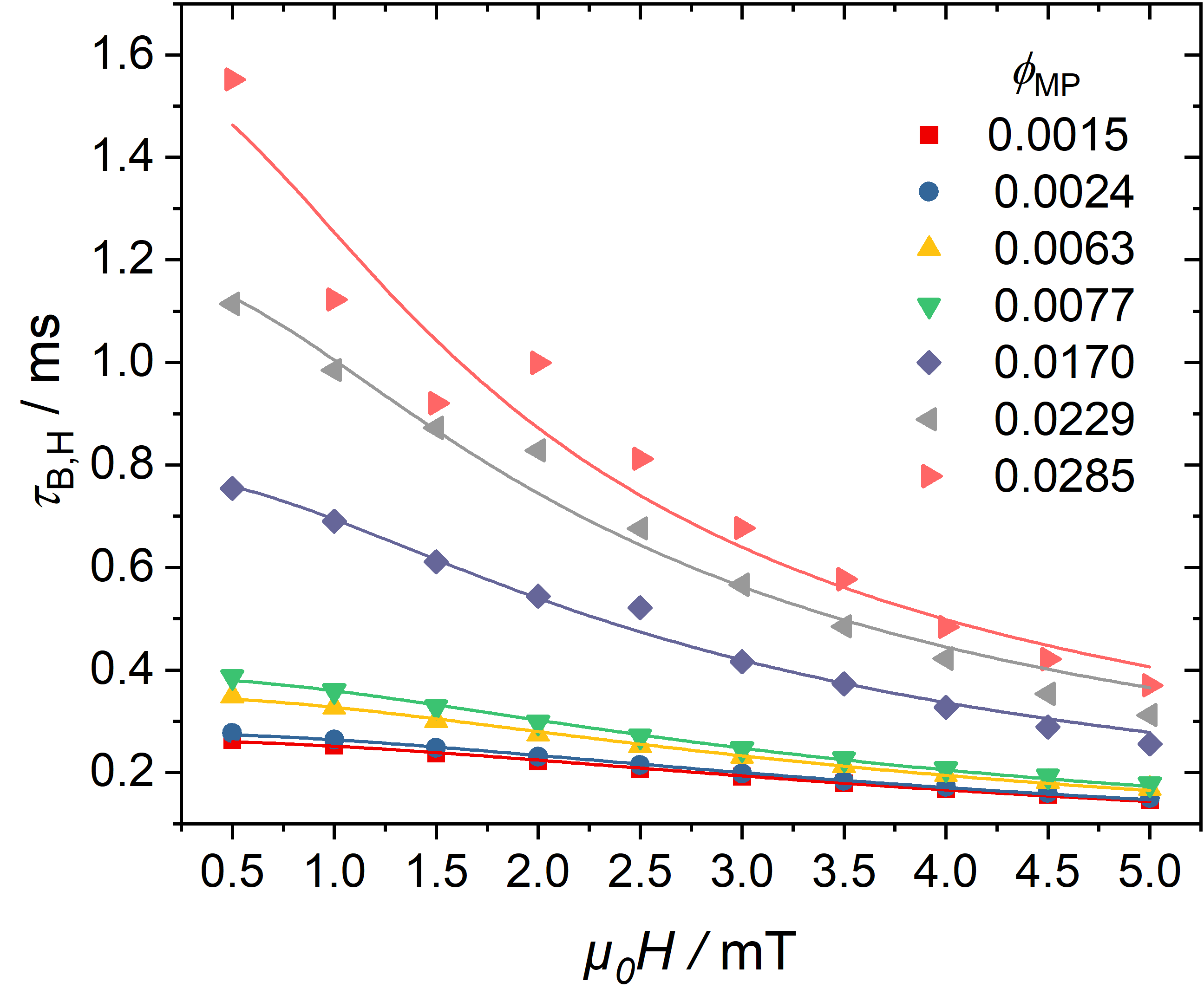}}
 \hspace{0.25em}
\subfloat[]{\includegraphics[height = 5 cm]{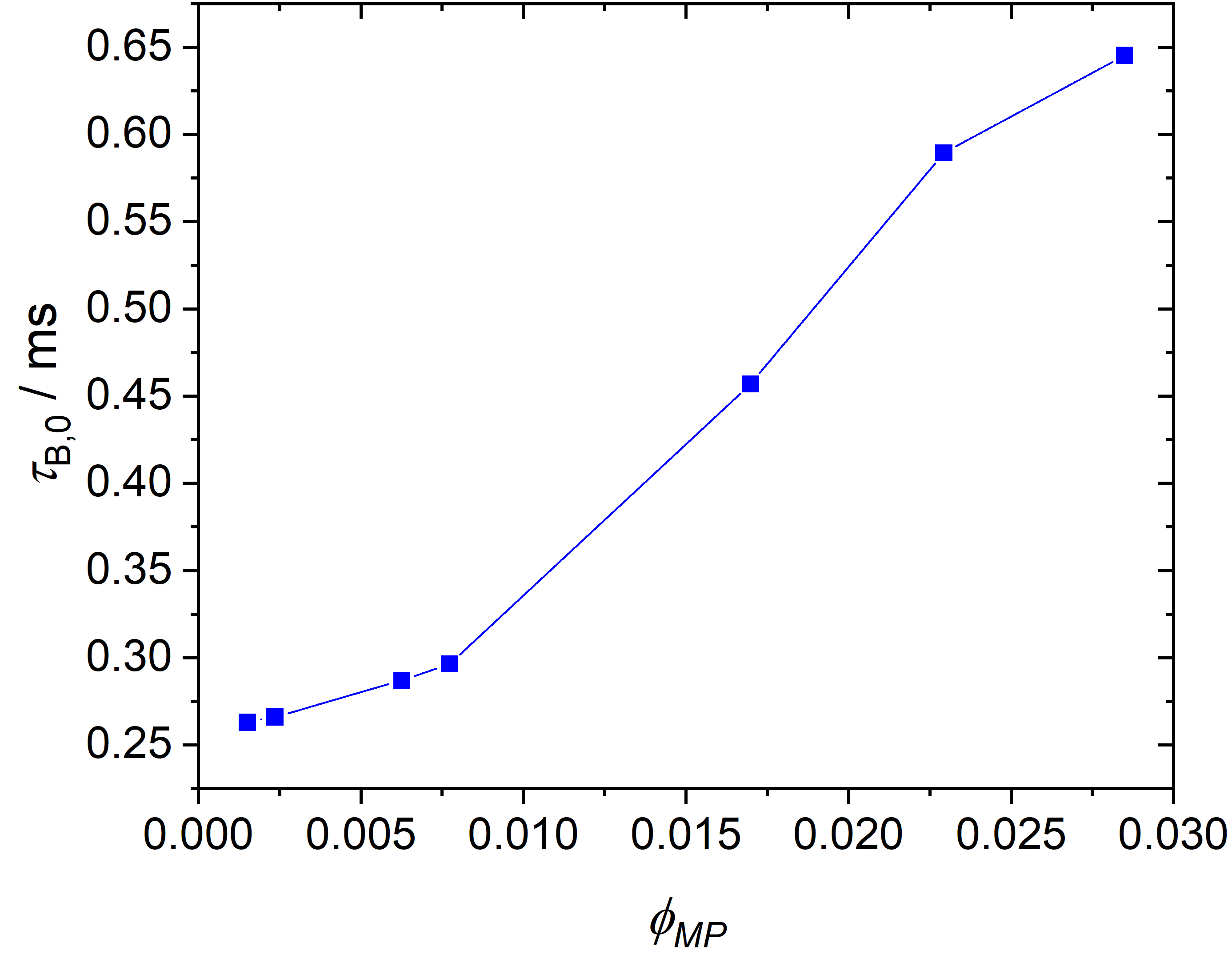}}
 %\hspace{0.25em}
\subfloat[]{\includegraphics[height = 5 cm]
{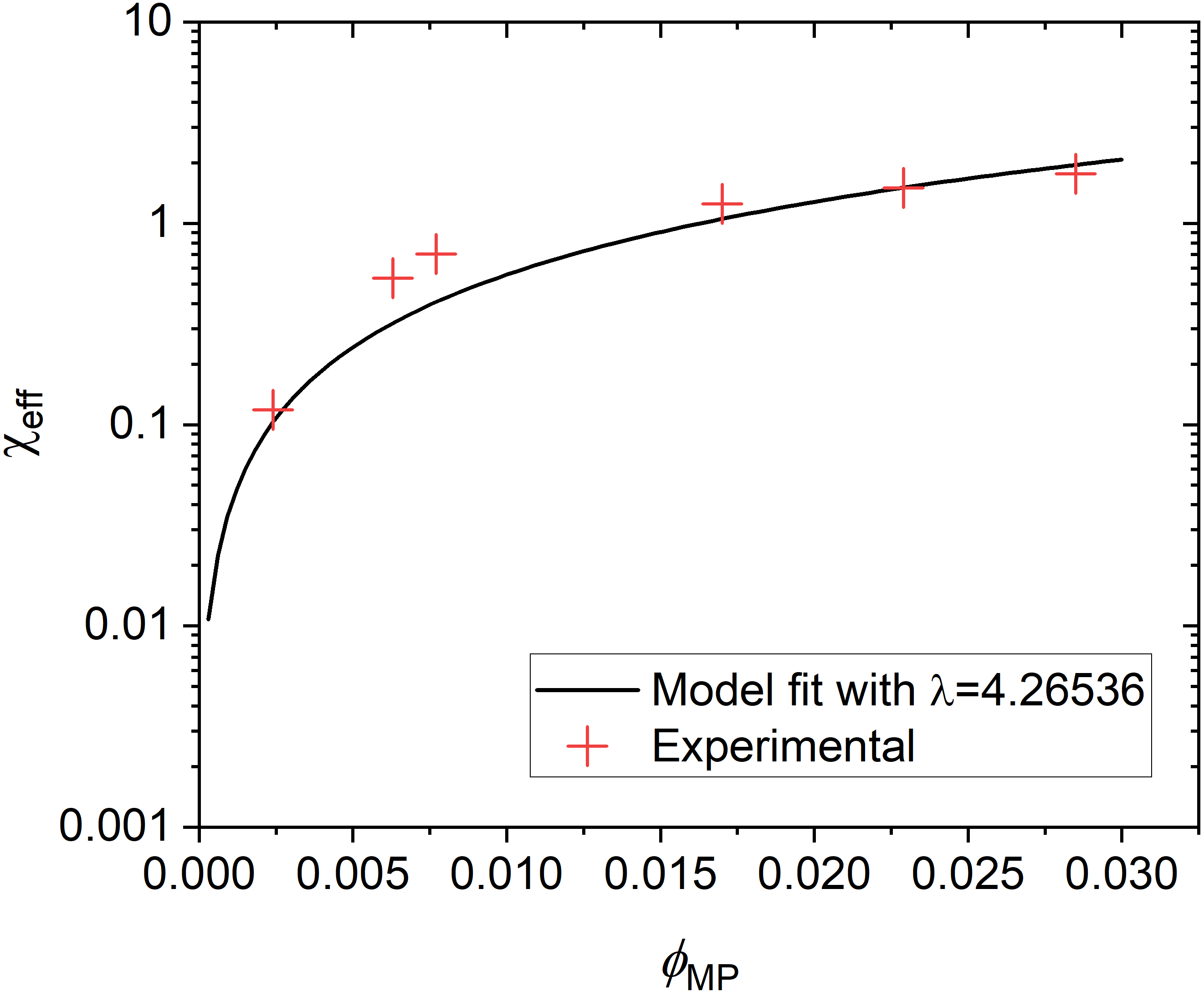}}
\hspace{1em}
\caption{(a) Field dependence of the Brownian relaxation time $\tau_{\mathrm{B,H}}$ for Set \textbf{2}. Lines show the fits with Eq.~(\ref{eq:Rusanov}). (b) shows the extracted field-free Brownian relaxation time $\tau_\mathrm{B,0}$ vs MP concentration $\phi_\mathrm{MP}$, with lines as guides to the eye. Figure (c) shows the effective static susceptibility $\chi_\mathrm{eff}$ vs the volume fraction $\phi_\mathrm{MP}$ (points) fitted to the chain-corrected model of Langevin susceptibility using the correct platelet partition function (line).}
\label{Rusanov}
\end{figure}

The field dependence of the Brownian relaxation time of individual platelets, as determined from the peak frequency of the HF mode, is depicted in Fig.~\ref{Rusanov}(a) in dependence of the applied field amplitude. The field-free Brownian relaxation time and its decay with increasing field amplitude strongly increase with $\phi_\mathrm{MP}$. The latter would reflect an increase in the effective coupling, which would be measured as an increase in the magnetic moment $m$.

Recently, Rusanov et al. \cite{Rusanov.2021} extended the model for the field-dependent Brownian relaxation time by Yoshida and Enpuku by additionally considering the effect of dipolar interactions. For the Brownian relaxations, they obtained the following empirical formula: 

\begin{equation}
\tau_{\mathrm{B,H}}=\frac{\tau_{\rm{B,0}}}{\sqrt{(1-\frac{\chi_\mathrm{eff}}{3})^2+0.076 \xi^2}}
\label{eq:Rusanov}
\end{equation}

Here the effective susceptibility $\chi_\mathrm{eff}$ describes the effect of magnetic interactions of the single platelets with the surrounding medium on the Brownian relaxation time $\tau_{\mathrm{B,H}}$. To fit the data in Fig.~\ref{Rusanov}(a), we determined the mean magnetic moment $m$ of the platelets and the field-free Brownian relaxation time $\tau_\mathrm{B,0}$ of noninteracting particles by fitting the data points of sample MP8 while setting $\chi_\mathrm{eff}=0$ (negligible dipolar interactions between particles). For the fits of the other data sets, $m$ was fixed at this value ($m=\qty{4.58e-18}{\ampere\meter\squared}$) while both $\chi_\mathrm{eff}$ and $\tau_\mathrm{B,0}$ were taken as free parameters (Fig.~\ref{Rusanov}). Although both the mean magnetic moment $m$ and the mean hydrodynamic size of the platelets are independent of MP concentration, the use of $\tau_\mathrm{B,0}$ as a free parameter is due to the fact that the viscosity is expected to increase with increasing $\phi_\mathrm{MP}$. The results for $\chi_\mathrm{eff}$ and $\tau_\mathrm{B,0}$ in dependence of $\phi_\mathrm{MP}$ are depicted in Fig.~\ref{Rusanov}(b) and \ref{Rusanov}(c). The viscosity increases by about a factor of 2, i.e., assuming that the viscosity of pure 1-butanol at room temperature amounts to \qty{2.54}{\milli\pascal\s}, it increases to about \qty{6.2}{\milli\pascal\s} for the sample with a Sc-BaHF concentration of \qty{158}{\gram\per\liter}. The rise of $\chi_\mathrm{eff}$ with $\phi_\mathrm{MP}$ is weaker than linear. 
The effective susceptibility $\chi_\mathrm{eff}$ in Eq.~\ref{eq:Rusanov} is dominated by clusters due to their high magnetic moments. The static susceptibility of clusters was theoretically described by Mendelev and Ivanov \cite{Mendelev:2004ve}. Adapting the partition function used in this theory to platelets (see SI), the expression for $\chi_\mathrm{eff}$ was fitted to the data points in Fig. 8(c). 
As a measure of magnetic dipolar interactions, the adjusted dipolar coupling constant $\lambda= \mu_0 m^2/(4 \pi D^3 k_{\mathrm{B}}T)$ is used, where $m$ is the magnetic moment and $D$ is the diameter of the platelets. In the curve in Fig.~\ref{Rusanov}(c), the aspect ratio of the platelet was taken to be $1:10$, which resulted in a fitted parameter $\lambda =4.3$. This aspect ratio corresponds to a mean thickness of 5 nm and diameter of 50 nm of the platelets as reported in \cite{Kuester.2022}. Note that a value $\lambda =4.1$ is calculated for $m=\qty{4.58e-18}{\ampere\meter\squared}$, $D=\qty{50}{\nano\meter}$ and $T=\qty{296}{\kelvin}$, in excellent agreement with the fitted value. Deviations of the fitted line from the experimental data in Fig.~\ref{Rusanov}(c) may be caused by the fact that the ratio between single and clustered platelets, which differently enter $\chi_\mathrm{eff}$, may vary when changing the concentration of Sc-BaHF, $\phi_\mathrm{MP}$. In addition, no electrostatic interactions but only dipolar ones are included, point-like dipoles rather than distributed and asymmetric ones are considered and distributions of parameters are not accounted for. 

\begin{figure}[hbt!]
\subfloat[]{\includegraphics[width = 0.45\columnwidth]{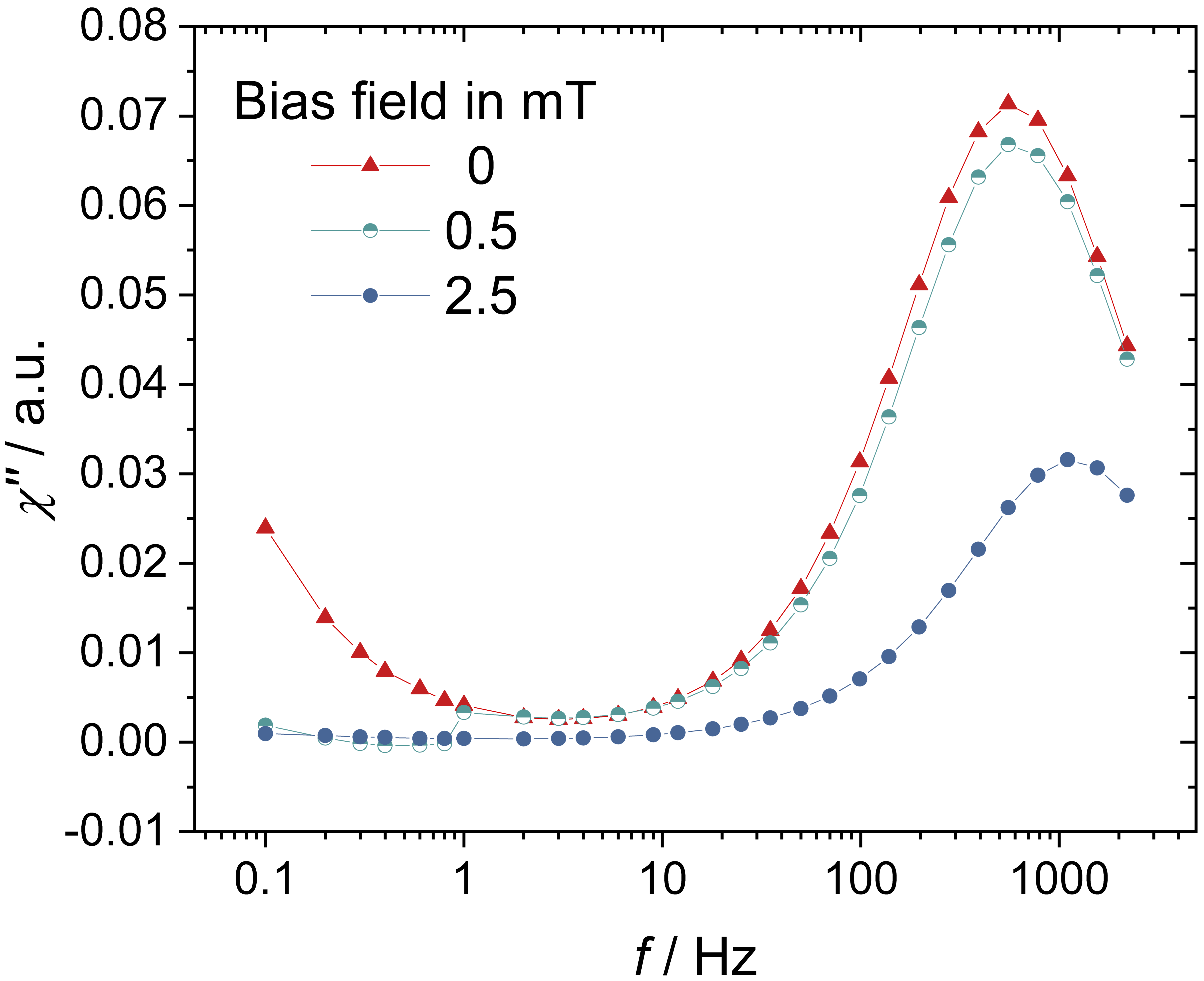}}
 \hspace{1em}
\subfloat[]{\includegraphics[width = 0.45\columnwidth]{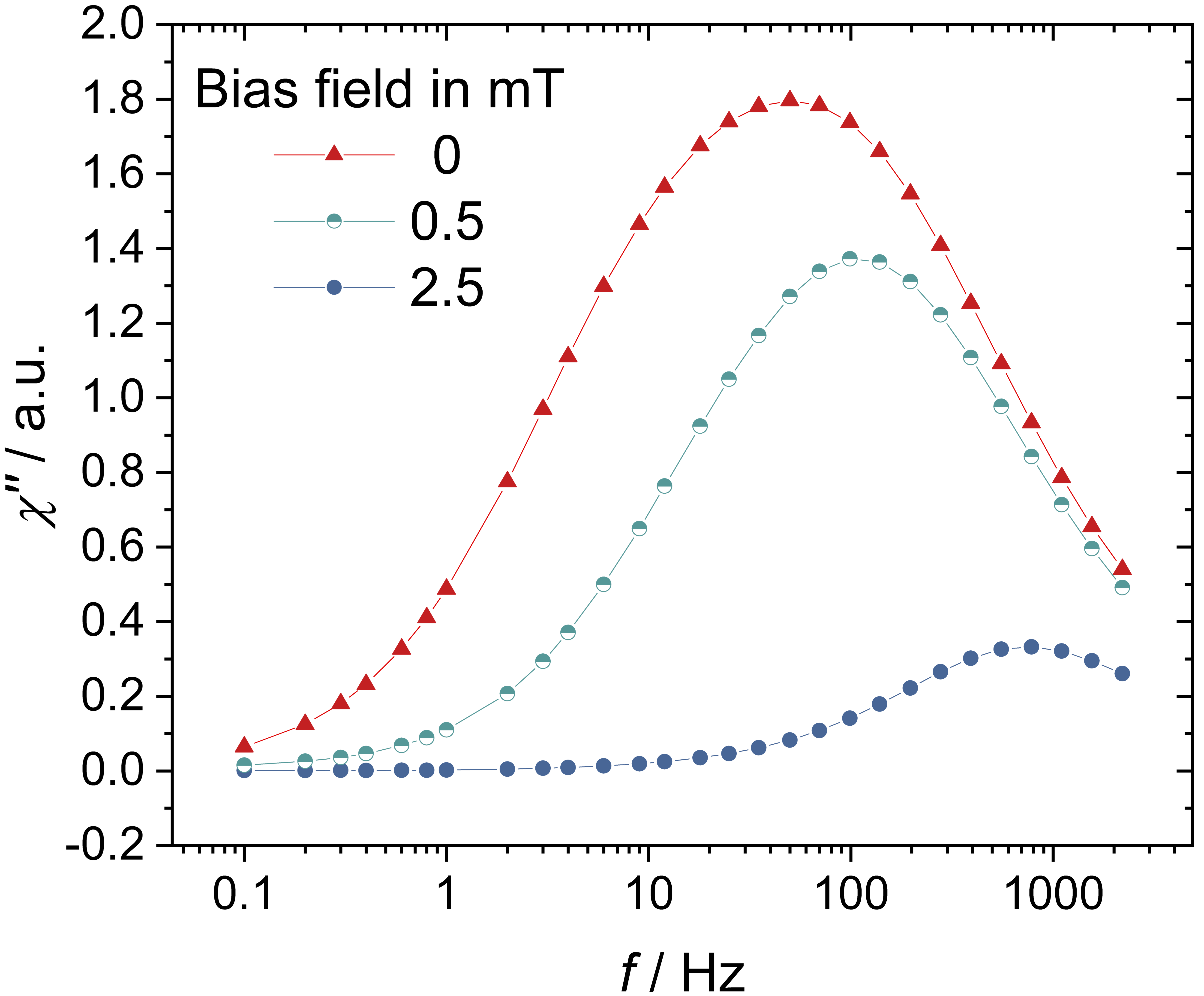}}

\caption{Influence of DC bias field applied parallel to the AC probe field on the ACS spectra of $\mathrm{MP8\_25}$  (a) and $\mathrm{MP158}$  (b). (a) In a low DC field (\qty{0.5}{\milli\tesla}), the low-frequency collective modes are suppressed as the clusters with high effective magnetic moment align first along the field. By a further increase of the DC field (\qty{2.5}{\milli\tesla}), the relaxation of single platelets with bigger size i.e. higher magnetic moment gets constrained, too. Hence, on (b), the single peak of the ACS spectrum belongs two more than one mode since the low DC bias field (\qty{0.5}{\milli\tesla}) mitigates it.}
\label{DCbias}
\end{figure}

\subsection{Effect of a DC-Bias Field}
Studying the effect of a DC bias field on the magnetic response allows us to selectively suppress the modes and explore their individual behaviour in dependence on the orientation of the bias field. 

One example containing pronounced LF and MF modes is $\mathrm{MP8\_25}$.
Applying a DC bias field parallel to the AC probe field for sample $\mathrm{MP8\_25}$ results in a suppression of the collective modes in the low-frequency range and, depending on the DC field strength, in a shift and repression of the HF peak maxima (Fig.~\ref{DCbias}(a)). The suppression of the low-frequency mode indicates, that the larger assemblies of magnetic nanoparticles are aligned with the field, and Brownian relaxation of the smaller particles now dominates the spectrum. Indeed, LF modes result from the collective behaviour of large ensembles (clusters) of MPs which 
are most susceptible to aligning along the DC field, hence as a primary response, a low (\qty{0.5}{\milli\tesla}) DC field suppresses the LF modes. 

Non-correlated single platelets couple individually to the magnetic field having much smaller coupling energy than the clusters do, and their relaxation can be quenched in significantly higher fields. As a result, the suppression of the HF peak occurs at a much higher DC bias field proving, that single platelets with comparably small $m$ contribute to the HF peak. The shift of the peak frequency to higher values and the decrease of its amplitude with increasing DC bias field strength are in agreement with theoretical models \cite{JETP:bp78JK4g,Coffey1992}.

This approach allows us to also characterise the magnetic dynamics in MP158 (Fig.~\ref{DCbias}(b)) having a single broad spectral feature in the ACS spectrum. With increasing DC bias field, the peak frequency shifts from about 50 Hz at zero bias field to 782 Hz at 2.5 mT. As seen in Fig.~\ref{DCbias}(a), a DC bias field suppresses the low-frequency part of the spectrum already at $H_{\mathrm{bias}}=\qty{0.5}{\milli\tesla}$ so that the peak at 2.5 mT is expected to be dominated by the HF mode. The fact that the peak frequency of the HF mode in Fig.~\ref{DCbias}(b) is slightly lower than that in Fig.~\ref{DCbias}(a) can be attributed to the effect of the Zeeman and dipolar interactions (cf. Sect. 3.2.1). For comparison, the peak position in the absence of bias field in Fig.~\ref{DCbias}(b) lies at 50 Hz for MP158, at a more than a magnitude lower frequency than that in Fig.~\ref{DCbias}(a) for MP8\_25 indicating, that the whole spectrum in Fig.~\ref{DCbias}(b) is the superposition of collective low-frequency and single-platelet high-frequency relaxation modes. 

\begin{figure}[h!]
  \includegraphics[width=\columnwidth ]{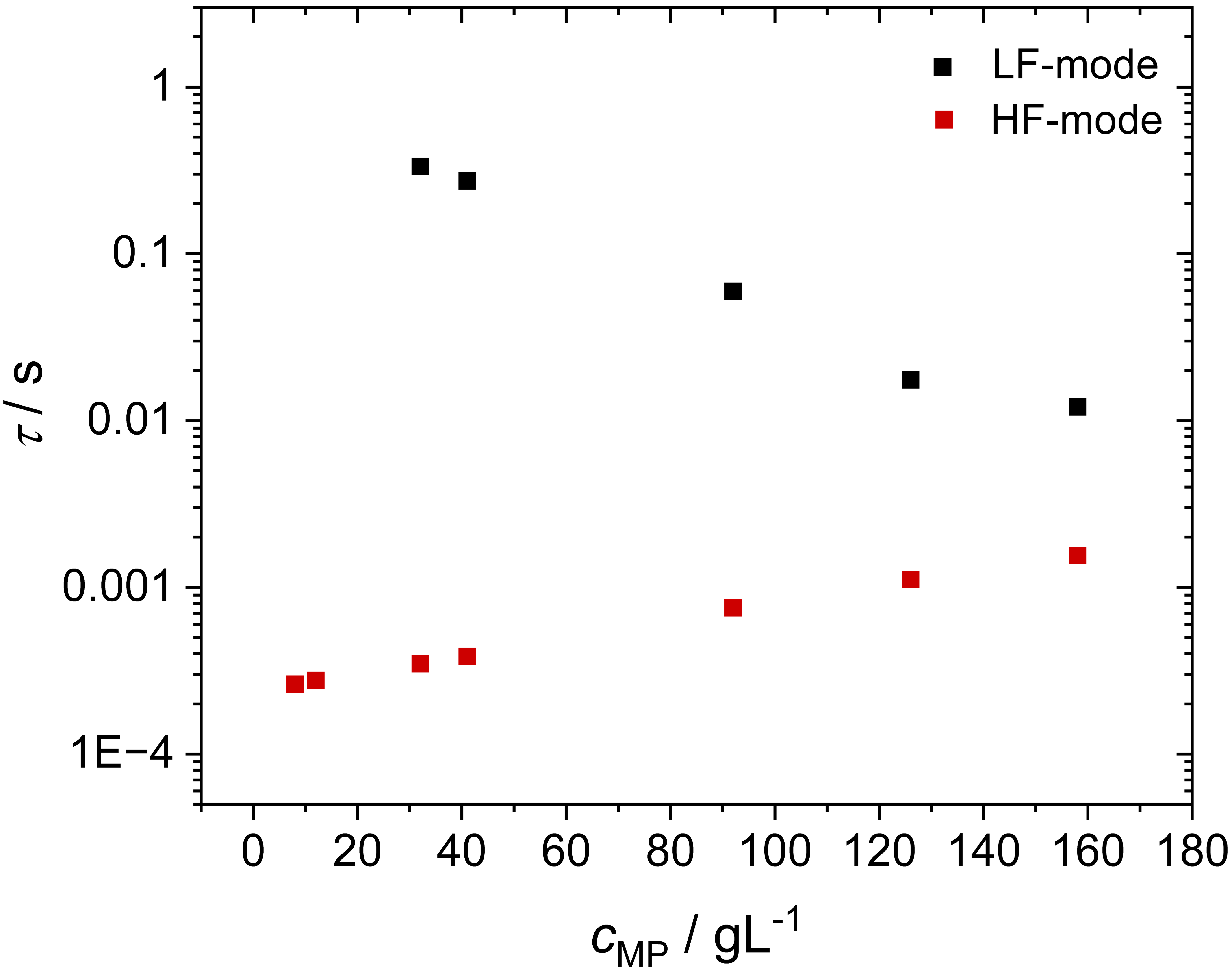}
  \caption{Relaxation times in a \qty{0.5}{\milli\tesla} DC bias field of the single-platelet (${\color{red}\smblksquare}$) and the collective (${\color{black}\smblksquare}$) modes as a function of the magnetic particle concentration for Set \textbf{2}. The two modes reveal opposite tendencies with increasing $c_\mathrm{MP}$: while the single-platelet mode slows down, the collective mode speeds up.}
    \label{fgr:tauLF_and_tauHF_vs_cMP}
\end{figure}

\section{Conclusions and outlook}
In our experiments we demonstrated that - tuning the electrostatic interactions by adjusting the DBSA concentration in the dispersions of Sc-BaHF nanoplatelets - strongly affects the structure of the low-frequency magnetic response (Set \textbf{1}). 
An increase in the DBSA concentration reduces the repulsive interactions and enhances interparticle correlations resulting in the growth of collective behaviour. Collective modes developing in the system are very sensitive to the magnetic field amplitude becoming faster in stronger fields, which suggests a high magnetic moment is associated with the relaxation modes. Another feature of these modes is their shift to higher frequencies with growing concentration of MPs (Set \textbf{2}). At the same time, the single-platelet relaxation mode becomes slower with increasing $c_\mathrm{MP}$, because of increasing dipolar interactions and increasing viscosity as it is expected (Fig.~\ref{fgr:tauLF_and_tauHF_vs_cMP}). As the MP concentration increases, the motion of the particles becomes strongly correlated leading to the development of orientational order and a restoring force, which accelerates the relaxation processes. This restoring force also contributes to the nematic order above a critical concentration ($c_\mathrm{MP}=\qty{126}{\gram\per\liter}$). At somewhat lower concentration, highly correlated clusters appear in the isotropic suspensions and are responsible for a strong magnetic and magnetooptical response. A similar situation of the so-called para-nematic state has been observed in dispersions of rod- and plate-shaped pigment particles~\cite{Eremin:2011th,May:2016ena, May.2017, May:2018fd}.

As a next step to determine the significance of viscosity change with increasing magnetic particle concentration, we are planning to measure viscosity of varied concentration of ferrofluids in and without magnetic field.
We will also elaborate on the ACS measurements in DC bias fields - parallel and perpendicular to the AC field - to comprehend more facets of the complexity of the spectra.

\section*{CRediT authorship contribution statement}
\textbf{Hajnalka Nádasi:} Conceptualisation, Investigation, Analysis, Writing - original draft, Writing - Review \& Editing.
\textbf{Melvin Küster:} Conceptualisation, Investigation, Analysis, Writing - original draft, Writing - Review \& Editing, Visualization.
\textbf{Alenka Mertelj:} Investigation, Resources, Writing - Review \& Editing.
\textbf{Nerea Sebastián:} Investigation, Resources, Writing - original draft, Writing - Review \& Editing.
\textbf{Patricija Hribar Boštjančič:} Investigation, Resources.
\textbf{Darja Lisjak:} Investigation, Resources.
\textbf{Thilo Viereck:} Conceptualisation, Resources, Review \& Editing.
\textbf{Margaret Rosenberg:} Computation, Simulations, Writing - Review \& Editing.
\textbf{Alexey O. Ivanov:} Conceptualisation, Computation, Writing - Review \& Editing.
\textbf{Sofia Kantorovich:} Conceptualisation, Computation, Writing - Review \& Editing.
\textbf{Alexey Eremin:} Conceptualisation, Writing - original draft, Writing - Review \& Editing.
\textbf{Frank Ludwig:} Conceptualisation, Investigation, Resources, Writing - original draft, Writing - Review \& Editing.

\section*{Conflicts of interest}
There are no conflicts to declare.

\section*{Acknowledgements}
F.L., M.K., A.E. and H.N. acknowledge the support of the Deutsche Forschungsgemeinschaft (Projects NA 1668/1-1 and LU 800/7-1). D.L, N.S., P.H.-B., and A.M. acknowledge the financial support from the Slovenian Research Agency (P1-0192, P2-0089, J1-2459, PR-08973 and PR-08415).
\appendix

%\bibliography{papers}
\bibliographystyle{elsarticle-num}


\end{document}